\documentclass[fleqn,usenatbib]{mnras}

\usepackage{newtxtext,newtxmath,lscape}

\usepackage[T1]{fontenc}
\usepackage{ae,aecompl}

\usepackage{graphicx}	
\usepackage{amsmath}	
\usepackage{multicol}

\title[On the edge]{On the Edge: the relation between stellar and dark matter haloes of Milky Way-mass galaxies}
\author[A. Genina et al.]{A. Genina$^{1}$\thanks{E-mail: agenina@mpa-garching.mpg.de (AG)}, A. J. Deason$^{2,3}$ and  C. S. Frenk$^{2}$  \\
$^{1}$ Max-Planck-Institut für Astrophysik, Karl-Schwarzschild-Str. 1, D-85748, Garching, Germany \\
$^{2}$ Institute for Computational Cosmology, Department of Physics, Durham University, South Road, Durham DH1 3LE, UK \\
$^{3}$ Centre for Extragalactic Astronomy, Department of Physics, Durham University, South Road, Durham DH1 3LE, UK
}

\date{Accepted XXX. Received YYY; in original form ZZZ}

\pubyear{2021}

\begin{document}
\label{firstpage}
\pagerange{\pageref{firstpage}--\pageref{lastpage}}
\maketitle

\begin{abstract}
  We investigate the build-up of the accreted stellar and dark matter haloes of
  Milky Way-like galaxies in APOSTLE suite of cosmological hydrodynamics
  simulations. We show that the stellar halo is made up primarily of
  stars stripped from a small number of massive dwarfs, most of which
  are disrupted by the present day. The dark matter halo, on the other
  hand, is made up primarily of small unresolved subhaloes
  ($\lesssim 10^6$ M$_{\odot}$) and a ``smooth'' component consisting
  of particles which were never bound to a subhalo. Despite these
  differences, the massive dwarfs that make up the majority of the
  stellar halo also contribute a significant fraction of the dark
  matter. The stars and dark matter stripped from these dwarfs are
  related through their kinematics and this leaves imprints in the
  phase-space structure of the haloes. We examine the relation between
  the location of features, such as caustics, in the phase space of
  the stars and dark halo properties. We show that the ``edge'' of the
  stellar halo is a probe of dark matter halo mass and assembly
  history. The edges of Milky Way-mass galaxies should be visible at a
  surface brightness of 31-36 mag arcsec$^{-2}$.

\end{abstract}




\section{Introduction}
\label{intro}

Within the Cold Dark Matter (CDM) model, structures form
hierarchically, with small haloes forming first and merging to
assemble larger objects \citep{fof}. In the early Universe, some of
the haloes have potential wells deep enough to allow the gas to become
cool and dense, resulting in the formation of the first stars. Smaller
haloes, with shallower potential wells may remain completely dark. The
formation of stars triggers the reionization of hydrogen and helium
atoms. The emitted radiation heats the halo gas above the virial
temperature, bringing star formation to a halt in small haloes. More
massive haloes will be able to maintain or re-accrete gas and continue
star formation over long periods of time \citep{efstathiou,thoul,bullock,benson, benitezfrenk}. These haloes host bright galaxies that we can observe today.

The process of hierarchical structure formation leads to the build-up
of the stellar and dark matter haloes of galaxies \citep{frenk85}. The
dark matter halo of our own Galaxy, the Milky Way, is thus predicted
to consist of its primordial dark matter component from the initial
peak collapse, smoothly accreted dark matter and the dark matter
that came from minor and major mergers with smaller haloes, some of
which have hosted stars \citep{eggen_lb_sandage, searle_zinn, white_rees, wang_aq} that now make up the stellar halo \citep{perek1951, roman1954}. Stellar haloes have also been identified in our nearest neighbour Andromeda \citep{mould_kristian, pritchet_vandenbergh} and beyond the Local Group \citep{davidge_pritchet, minniti_1996, minniti1999, harris_harris, sarajedini, chapman_2006, ibata_2007, durrell_2010, monachesi_2013, harmsen_2017}.

The majority of the Milky Way's stars belong to its disk component and the
bar \citep{milkywaymass}; however a non-negligible fraction resides in the Milky Way's
extended stellar halo. Within the CDM paradigm,
these halo stars are the result of tidal stripping of infalling dwarf
galaxies. Some fraction of halo stars could also have come
  from the heating of the stellar disk stars by supernovae feedback or
  molecular clouds, or from encounters with dwarf galaxies
  \citep{bensona,zolotov_halo_insitu, font_insitu, yu_outflow_halo,
    gomez}. A fraction of the stellar halo mass also stems from accreted globular clusters \citep{searle_zinn}, some of which can be observed in the process of disruption today \citep{odenkirchen_2001, mateu_2018, starkman_2020, piatti_2020}.  In this work, we exclusively consider the build up of the stellar haloes due to accretion.  The Sagittarius dwarf galaxy
\citep{ibata_sag1,ibata_sag2} and its extended tidal tails
\citep{majewski_sag_tails} are an example of this process happening at
the present day. Past mergers experienced by the Milky Way are
expected to have left an imprint in the phase-space structure of our
Galaxy. Stars coming from the same progenitor can be seen to ``clump''
in angular momentum -- energy space \citep{helmi_de_zeeuw,gomez_helmi}. Spatially, mergers can leave an imprint in the form of
`shells' and `streams' which have been observed in the Milky Way and
nearby galaxies \citep{quinn, ibata_m31, mcconnachie_m31,cooper_shell, bernard_panstarss,shipp_des, stellar_stream_legacy_survey}. The
identification of these features in the spatial distribution and
kinematics of Milky Way's stars, coupled with their chemical
abundances, give clues to the accretion history of our Galaxy
\citep{johnston_accretion_history, bonaca}. Recently, some of these
properties have been used to determine that Milky Way underwent a
merger with an object now known as {\it Gaia Enceladus/Sausage}
\citep{helmi_enceladus,sausage}. Other progenitors of the present-day
Milky Way have been inferred from the chemo-kinematics of the Milky Way's
accreted population of globular clusters \citep{kruijssen_gc}.

In recent years, the {\it Gaia} satellite \citep{helmiGaia, gaia_edr3}
has uncovered  a number of disrupted objects within the
Milky Way through the coherent kinematics of their stripped
stars. Tools such as {\sc streamfinder} \citep{streamfinder} have been used to discover
such objects with proper motions and deep photometry. This approach has been
effective at identifying a number of globular cluster streams within
the Milky Way. Some of these have been associated with dwarf galaxies
which now make up the stellar halo
\citep{malhan_mw_mergers}. Nevertheless, finding evidence
of disrupted dwarf galaxies in stellar motions has proven to be more
difficult. This is because the large velocity dispersion in dark
matter-dominated dwarf galaxies result in more kinematically hot
streams, where the orbits of stripped stars can vary substantially
from that of the dwarf \citep{helmi_white}. Moreover, due to their
higher mass, dwarf galaxies tend to sink into the centre of the Galaxy
by dynamical friction \citep{amorisco_halo}. The stripped stars,
particularly near the centre, become phase-mixed over time and the
effectiveness of the integrals of motion in identifying coherent
structures is then limited in the time-varying asymmetric potential of the
Milky Way. For these reasons, it is the stars in the outer halo, with
longer dynamical times, that likely hold clues on past mergers
of our Galaxy.

The CDM paradigm, where the halo assembles largely through tidal
stripping of smaller infalling objects, predicts that the majority of
the stellar halo and at least a fraction of the dark matter halo have
a common origin. It is thus possible that the properties of the Milky
Way's stellar halo can be used to investigate those of the dark
matter. For instance, the extent of the stellar halo may be directly
related to that of the dark matter. Moreover, phase-space features in
the stars may also suggest equivalent features in the dark
matter \citep{tissera_2014, herzog-arbeitman}. Local dark matter overdensities are important features for 
direct and indirect searches for the dark matter particle
\citep{simpson_aurigaia,necib_gaia, darkshards}.

In the spherical collapse model of the formation of
virialized structures, overdensities in the early Universe
gravitationally attract surrounding material, causing it to collapse
and virialize, leading to the formation of `caustic' shells of matter \citep{vogelsberger_caustic1, vogelsberger_caustic2},
corresponding to the apocentres of successively accreted material,
with their spacing dependent on the rate of growth of the dark matter
halo \citep{gunn_gott,fillmore, bertschinger}. Although this picture
is simplified, structures modeled with $N$-body simulations match well
analytical predictions \citep{zavala, adhikari, sugiura}. In particular,
the outermost shell, corresponding to the first apocentre of the most
recently accreted material is related to the ``splashback'' radius of the
halo and provides a physical definition of the halo boundary
\citep{diemer_and_kravtsov, diemer_2017}. With the inclusion of
hydrodynamical processes in cosmological simulations, it has become
possible to follow the evolution of the stars and gas after
accretion. \cite{edgeofthegalaxy} have shown that together with the
outermost `splashback' radius, Milky Way analogues in the APOSTLE
\citep{sawalapuzzles,fattahi}, AURIGA \citep{auriga} and ELVIS
\citep{elvis} simulations also have a `second caustic' in the dark
matter, which roughly coincides with the visual extent of the stellar
halo and is located at $\sim 0.6 R_{200,\rm m}$, or near
$R_{200, \rm crit}$. This second caustic is directly measurable as the
steepest drop in the log-slope of the stellar density distribution or
in its radial velocity profile. An observation of the "edge" feature
could thus allow us to infer the size of the dark matter halo and its
properties would be directly related to the accretion history of the
Milky Way.

In this work, we investigate the origin of the `second caustic' in the
stellar and dark matter haloes of Milky Way~/~M31 analogues in APOSTLE
simulations. In
Section~\ref{simulations}, we introduce the APOSTLE suite of
simulations and provide some definitions that we will use throughout
this work. We then split our sample of 10 galaxies into a `quiet'
subsample that is more Milky Way-like and an `active' subsample that
is more M31-like \citep{deason_mw_m31,pillepich,lachlan}. In
Section~\ref{buildup} we examine the historical build-up and
present-day composition of the stellar and dark matter haloes of these
analogue galaxies. In Section~\ref{phasespace}, we examine the
phase-space properties of the `quiet' and `active' galaxies, focusing
on the differences between the two and the relation to the phase-space
properties of their dark matter haloes. In
Section~\ref{formationofcaustics}, we investigate the formation of
phase-space features in the stellar and dark matter haloes, looking in
particular at the mergers that contributed to their formation, their
infall and tidal history. We comment on the halo and galaxy properties
which influence the characteristics of the phase-space
distribution. In Section~\ref{observations}, we discuss how
observations of the luminous stellar component of haloes may be used
to uncover the properties of the dark component and observational
strategies that the present work suggests. In
Section~\ref{conclusions}, we summarise our results.

\section{Simulations}
\label{simulations}

\subsection{APOSTLE simulations}

The APOSTLE (A Project Of Simulating The Local Environment)
simulations are a suite of $N$-body hydrodynamical zoom simulations of
environments resembling the Local Group
\citep{fattahi,sawalapuzzles}. Each simulation volume features a Milky
Way - M31 analogue pair. The pairs of haloes were selected to match
the observational constraints on the Local Group, such as the combined
halo mass, galaxy separation, relative radial and tangential velocity
and the velocities of nearby galaxies. The M$_{200}$ values of the
haloes range between $5\times10^{11}$ and
$2.5\times10^{12}$~M$_{\odot}$. The high-resolution zoom region
comprises a sphere of $\sim2.5$~Mpc from the barycentre of the halo
pair, within a $100^3$~Mpc$^3$ box. The suite consists of 12 volumes
simulated at low and medium resolution, while 5 volumes have also been
simulated at high resolution. In this work, we analyse these five
simulations, giving us a sample of 10 Milky Way-mass galaxies.  Their
dark matter particle mass is in the range
$2.5 - 5\times10^4$M$_{\odot}$ and their gas particles have initial
masses in the range $0.5-1\times10^4$M$_{\odot}$; the gravitational
softening length, $\epsilon_g = 134$~pc.

The APOSTLE suite was run with the {\sc p-gadget-3} code
\citep{gadget}, assuming a WMAP-7 cosmology \citep{wmap7}. A {\sc
  Tree-PM} scheme is used to compute gravitational
accelerations. Galaxy formation is modeled using the {\sc eagle} code
\citep{eagle1,eagle2}. {\sc Eagle} solves hydrodynamic forces using
the smoothed particle hydrodynamics (SPH) {\sc Anarchy} scheme
\citep{anarchy,anarchy2} and the pressure-entropy formalism
\citep{hopkins}. The {\sc eagle} model was calibrated to reproduce the
$z=0.1$ stellar mass function and galaxy sizes above
$10^8$~M$_{\odot}$. The model includes cooling, star formation and
evolution and feedback from supernovae, stellar mass loss, active
galactic nuclei and radiation pressure
\citep{gascooling,starformation,starformation2,agnbooth,accretionmergers}. A
uniform ionizing background is turned on instantaneously at $z=11.5$
\citep{haardtmadau}. Cooling rates are computed for 11 tracked
chemical elements (including hydrogen and iron), assuming ionization
equilibrium in the presence of UV and X-ray backgrounds and the cosmic
microwave background \citep{enrichment}.

Artificial fragmentation of the ISM is prevented by imposing a
temperature floor through a polytropic equation of state. Star
formation has a metallicity-dependent density threshold that
effectively ranges between $n_H$ = $0.1-1$~cm$^{-3}$. Star formation
is also pressure-dependent and follows the Kennicutt-Schmidt star
formation law \citep{ks1, ks2}. A stellar particle within the simulations represents a
simple stellar population following a \cite{chabrier} initial mass
function. Feedback from star formation is implemented using the
stochastic thermal prescription of \cite{anarchy}. 

{\sc Eagle} has been shown to reproduce the evolution of the stellar
mass function, colours and magnitudes of galaxies, scaling laws of
galaxy populations. The {\sc eagle-ref} model, used in APOSTLE, produces Milky Way and M31 analogues of lower stellar mass than suggested by observations, as discussed in \citet{fattahi}. At the same time, APOSTLE provides a good match to the
abundances of satellites and dwarf galaxy scaling relations within the
Local Group \citep{sawalapuzzles, fattahi, campbell}, including dwarf metallicities
\citep{thedistinct}.

\subsection{Classification of stellar and dark halo components}

In this work we investigate the build-up of the accreted stellar and dark
matter haloes of Milky Way-mass galaxies. In order to do so, we must
define a redshift-dependent halo boundary. A possible definition of
this is the splashback radius; however, the exact definition of this
radius is uncertain and complicated by the fact that we are studying
group environments. Instead, we opt to use the radius enclosing 200
times the mean matter density of the Universe at each redshift,
$R_{\rm 200,m}(z)$, as our definition of the halo boundary. This is
also motivated by the finding that the splashback radius tends to be
close to this value, and much further our than $R_{200,\rm crit}$
\citep{diemer_and_kravtsov}.  We identify substructures in our
simulations using the {\sc HBT+} algorithm \citep{hbt}. For each
stellar particle within $R_{\rm 200,m}(z)$ of either of the main
haloes at $z=0$, excluding those that are identified as bound to
satellite haloes, we track the particle back in time until it is
identified as being bound to a halo or a subhalo that is not one of
the two main haloes.  We classify all stellar particles which are
bound to the Milky Way~/~M31 analogue at the time of their birth as
`disk' particles. A negligible fraction of stellar particles are never
identified as bound in the available snapshot outputs and we exclude
these particles from the analysis.

For the dark matter halo particles, we also track each particle back
in time until there is a substructure match. However, unlike the
stellar particles, a large fraction of dark matter comes from smooth
accretion. This includes unresolved haloes and individual dark matter
particles that were never bound to a subhalo. All dark matter
particles which are part of one of the main haloes at $z=0$ and for
which there is no historical substructure match are classified as the
`smooth' dark halo component.

We also must decide how to treat mergers of substructures which
occur {\it inside} of the Milky Way/M31 halo. We opt to count the
subhaloes that merged while within the halo as a single object, even
if particles had been stripped within the Milky Way/M31 analogue prior
to the merger. Additionally, we impose a merging criterion. Namely, we
consider a subhalo to have merged with another subhalo when it has
lost all of its bound mass, retaining only an `orphan' particle. To be
merged with a subhalo, this particle must lie within the maximum
subhalo radius, defined by the location of subhalo's furthest bound
particle. We further ensure that this particle is gravitationally
bound to the subhalo. We establish whether the orphan particle is
bound by comparing its relative velocity with respect to the centre of
the subhalo to the subhalo's local escape velocity. The latter is
computed by performing an NFW profile fit to the subhalo out to the
maximum subhalo radius. Finally, at $z=0$, we `climb the tree' and
ensure all orphans representing substructures and
substructures-of-substructures are assigned to the main progenitor.

Finally, due to the time spacing of outputs of our simulations some stellar particles may be incorrectly identified as formed {\it in-situ} within the Milky Way/M31 analogues if the star-forming gas of an accreted satellite has been stripped between two simulation snapshots \citep{zolotov_halo_insitu}. To correct for this, we take all stars that are identified as {\it in-situ} at birth and check if their parent gas particle was bound to the Milky Way/M31 analogue or a satellite galaxy in the previous snapshot and assign its origin accordingly.

\section{The build-up of the accreted stellar and dark matter haloes}
\label{buildup}

\begin{figure*}
    \centering
    \begin{multicols}{2}
    \includegraphics[width = \columnwidth]{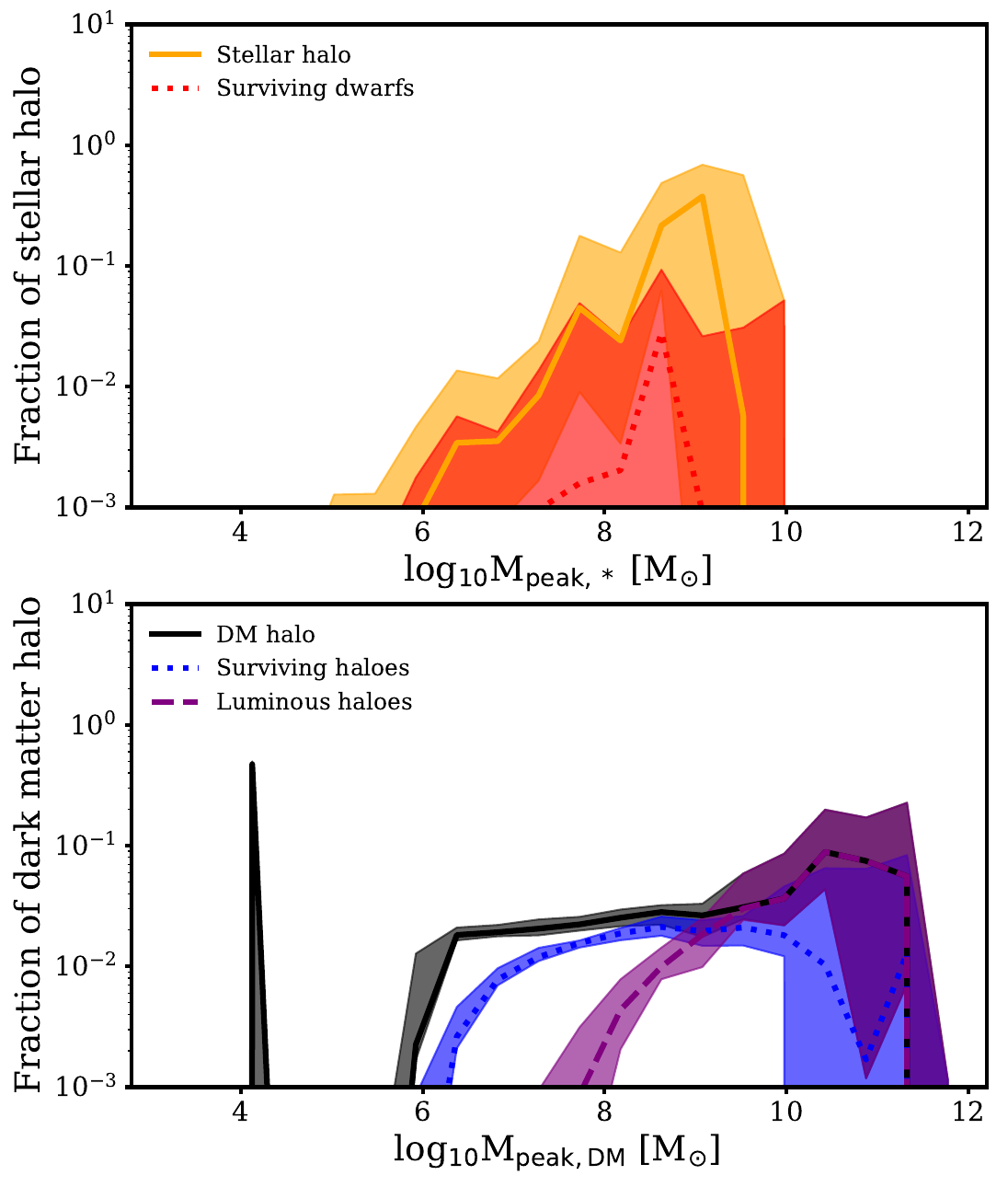}
    \includegraphics[width = \columnwidth]{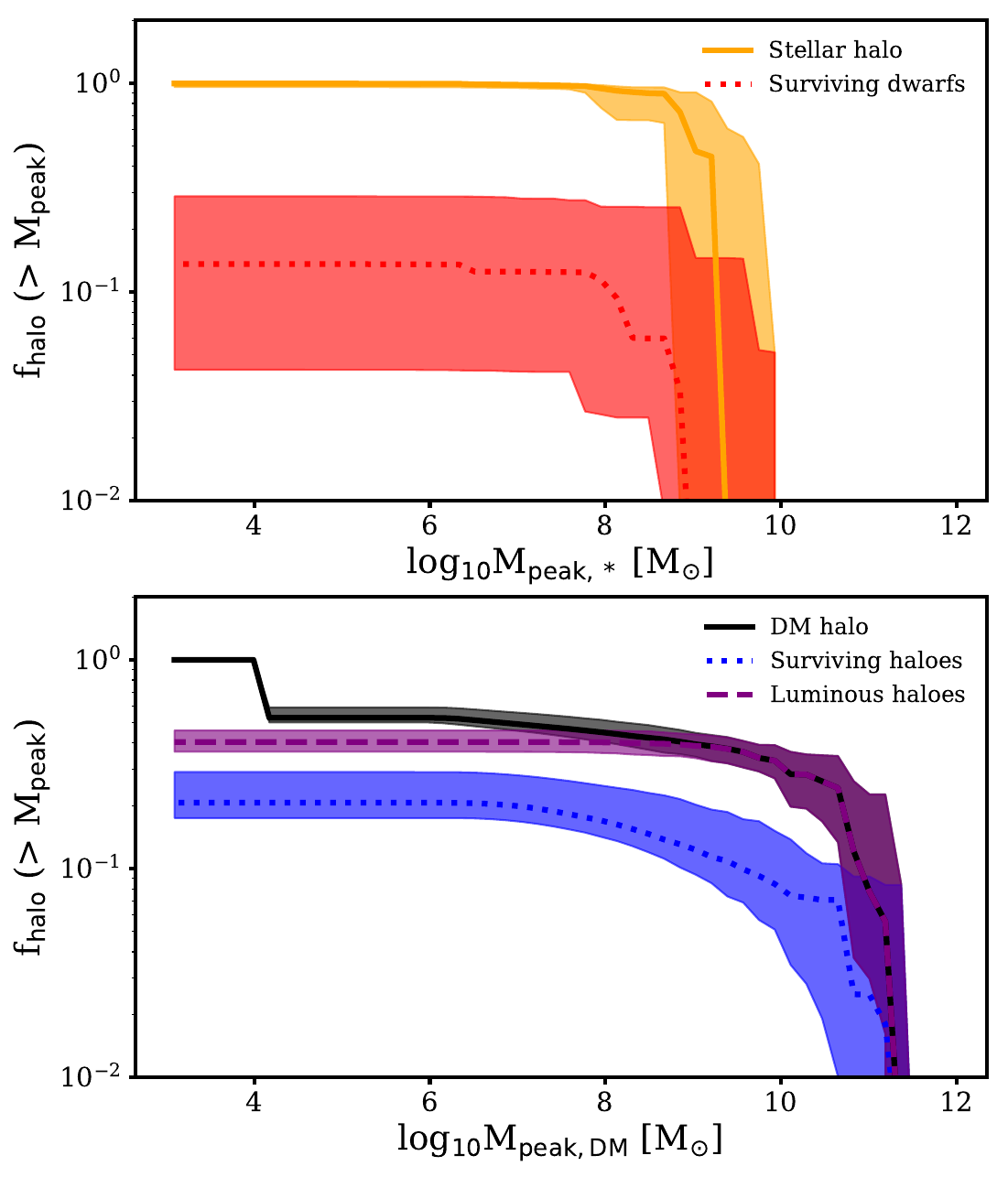}\\
    \includegraphics[width = \columnwidth]{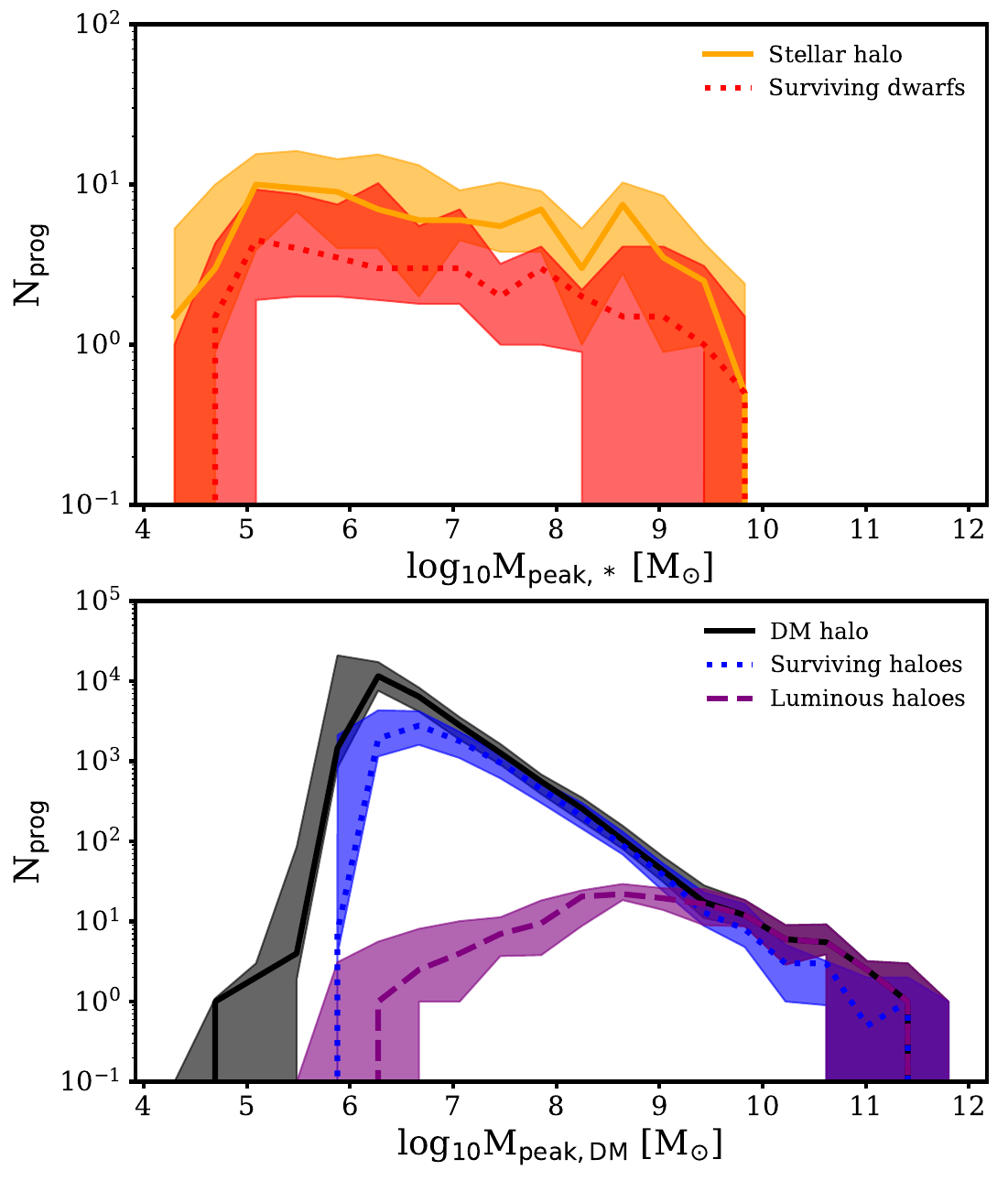}
    \includegraphics[width = \columnwidth]{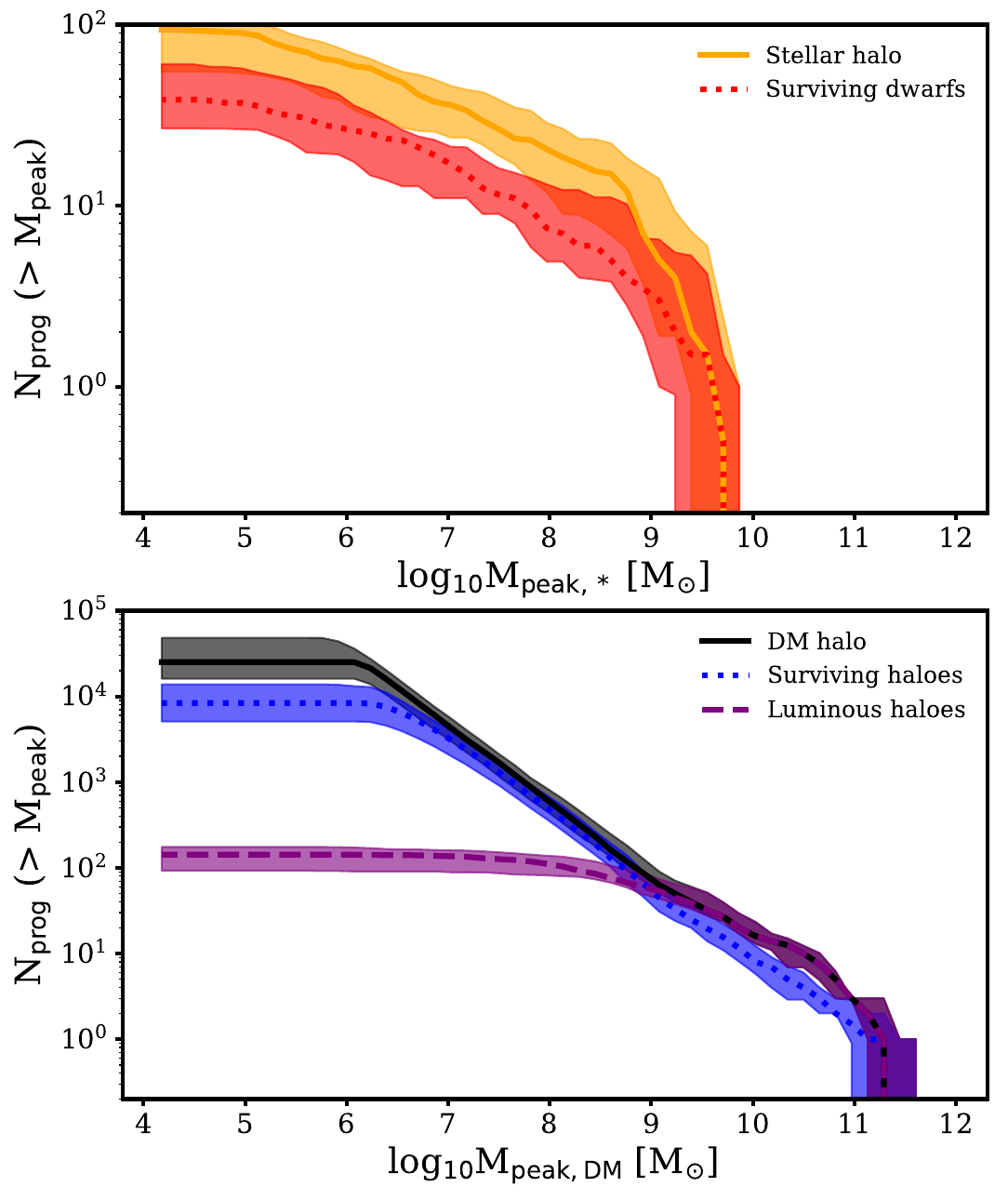}
    \end{multicols}
    \caption{Composition of the stellar and dark matter haloes of
      Milky Way-M31 analogues, defined by the stars and dark matter
      that are not bound to any subhalo within R$_{200, \rm m}$. \textit{Top
        left:} the fraction of stellar and dark matter haloes made up
      from progenitors of a given peak stellar and dark matter masses.  Thick
      lines show the median and the bands span the minimum and
      maximum values of the entire sample. Yellow corresponds to the
      make-up of the entire stellar halo and red identifies the
      objects that survive  to the present day. Black shows the
      entire dark matter halo, purple the subhaloes that
      have hosted stars in the past and blue the subhaloes
      that survive  to the present day. The peak near $\sim10^4M_{\odot}$
      corresponds to the typical mass of a dark matter particle in
      APOSTLE and shows the contribution of the smooth
      component. \textit{Top right:} halo composition in terms of the
      number of contributing objects of each peak stellar/halo
      mass. \textit{Bottom left:} the fraction of the halo made up of
      objects above a given peak mass. \textit{Bottom right:} number of
      dwarfs/subhaloes above a given peak mass contributing stars and
      dark matter to the halo.}
    
    \label{fig1}
\end{figure*}

\begin{figure}
    \centering
    \includegraphics[width =\columnwidth]{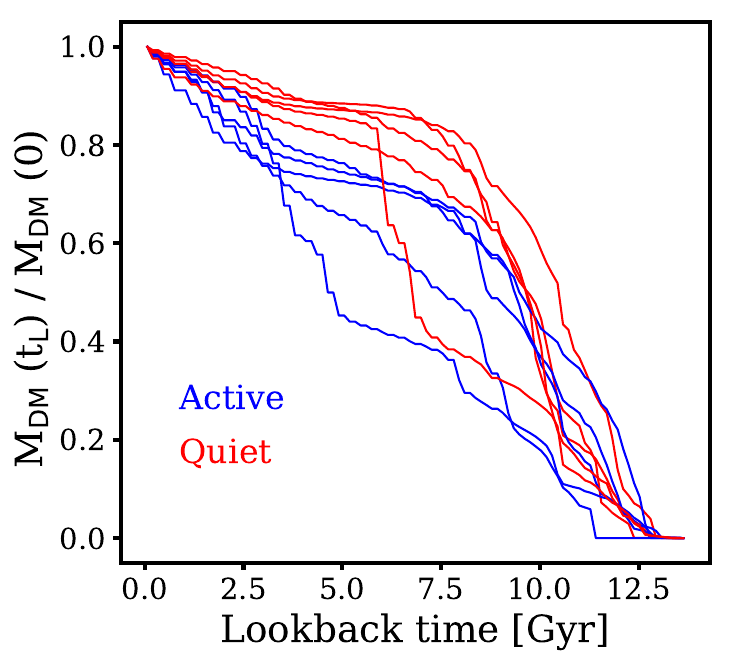}
    \caption{The dark matter assembly histories of the Milky Way/M31
      analogues in our simulations. The haloes are split into
      ``active'' and ``quiet'' based on their recent accretion history
      (within the last 6~Gyr), namely ``quiet'' haloes that were assembled
      early and ``active'' haloes that show fast recent growth. }
    \label{fig2}
\end{figure}

\begin{figure*}
    \centering
    \begin{multicols}{3}
    \includegraphics[width = 1.1\columnwidth]{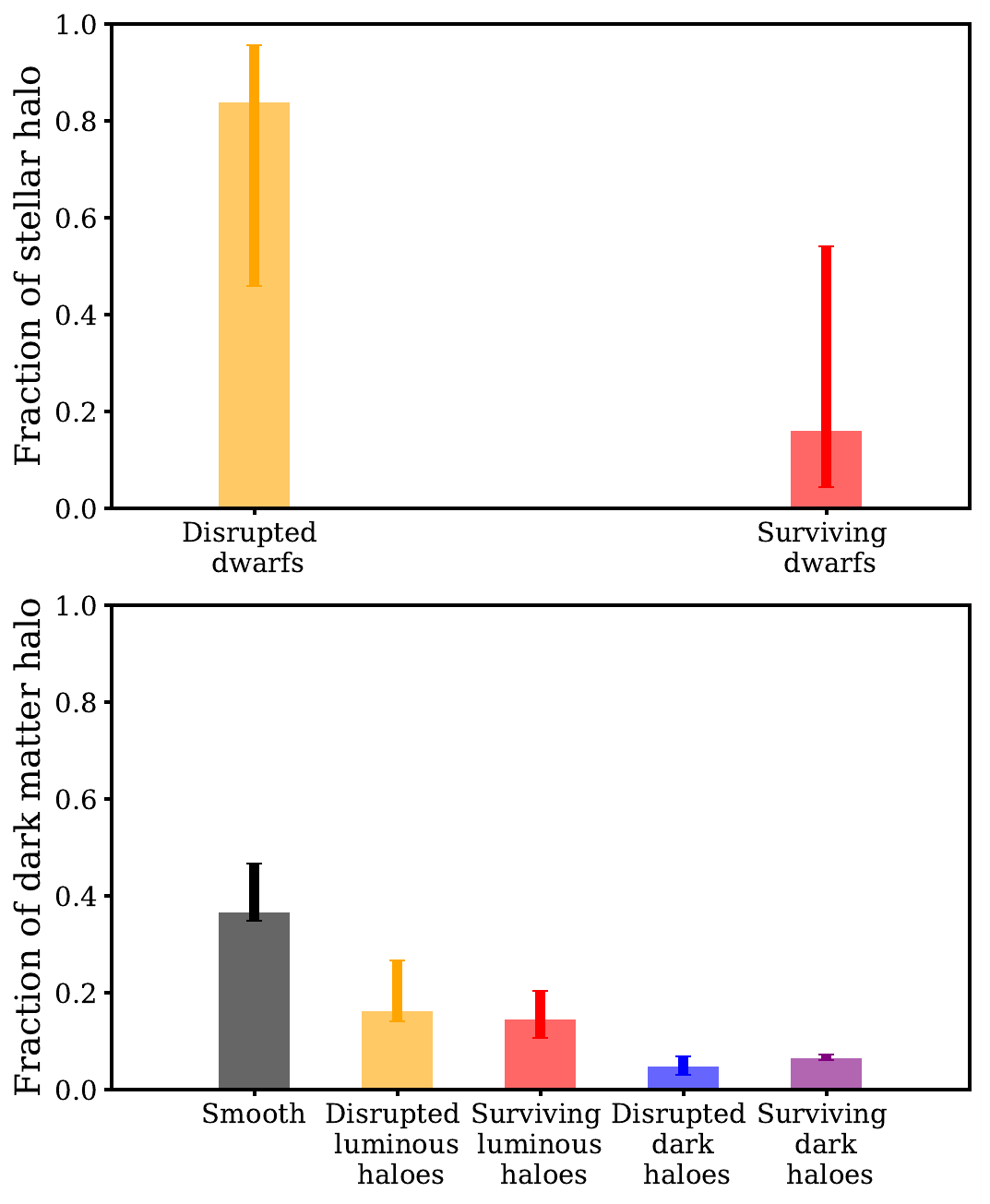}

    \includegraphics[width = 1.1\columnwidth]{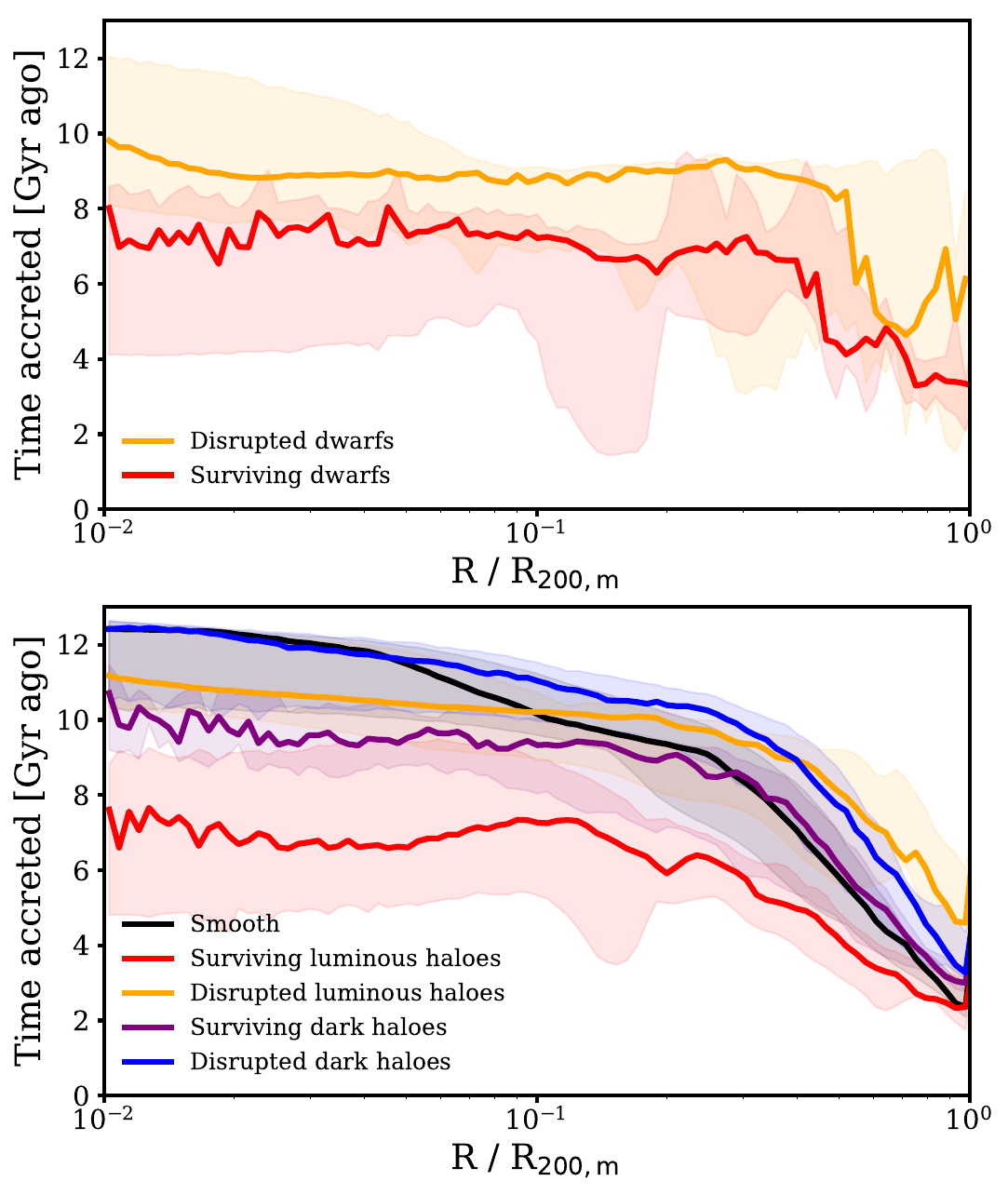}\\

     \includegraphics[width = 1.1\columnwidth]{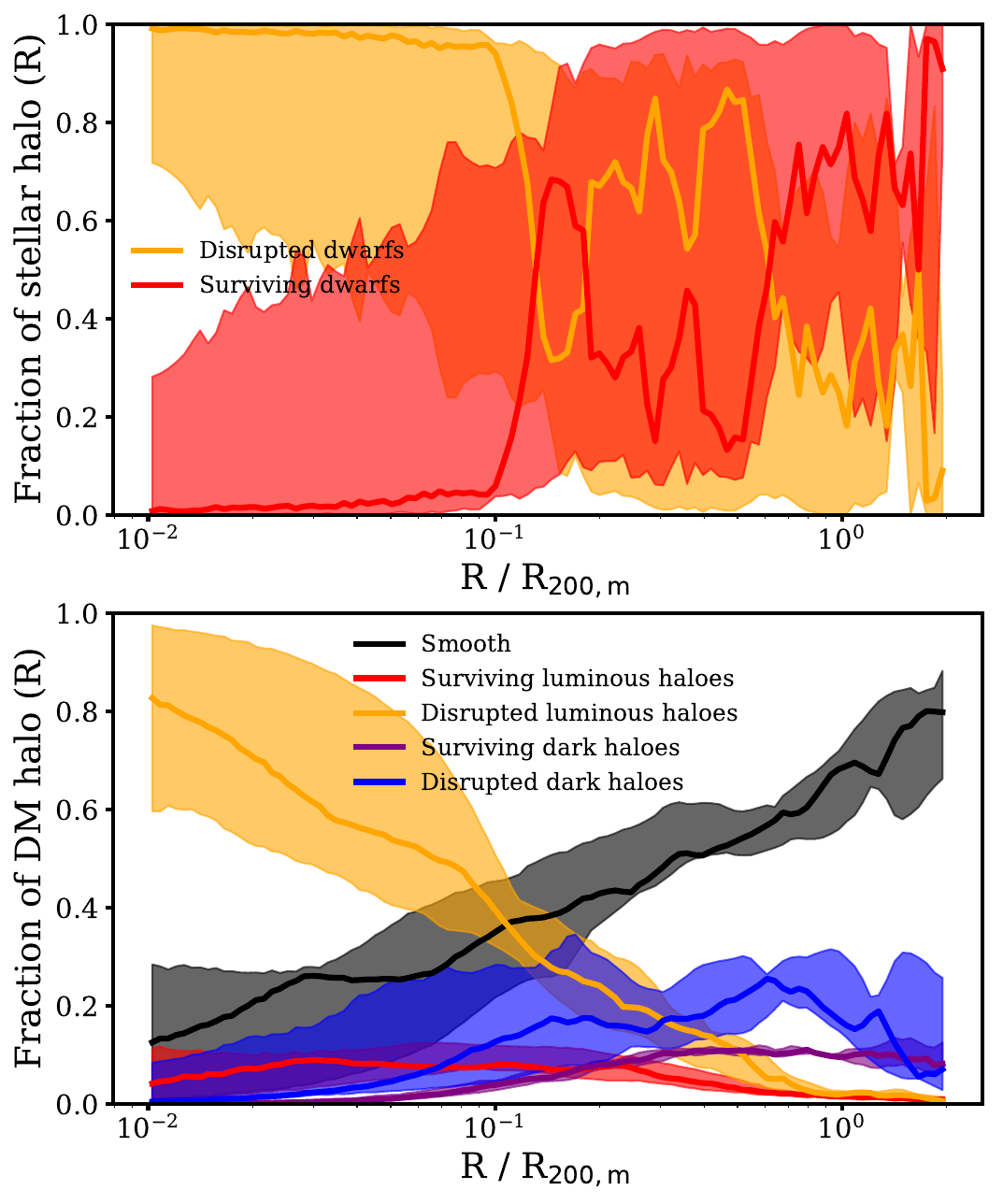}

    \includegraphics[width = 1.1\columnwidth]{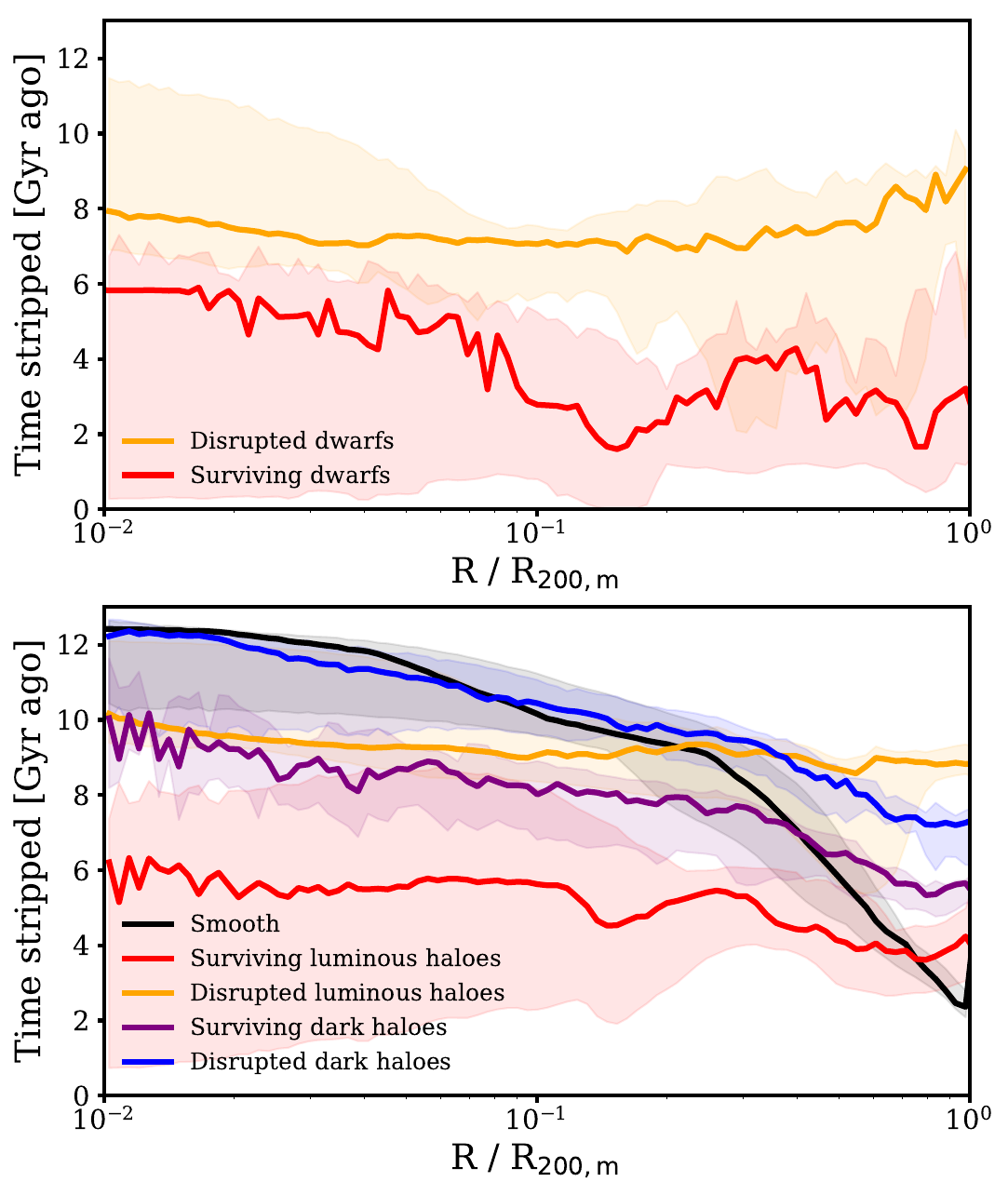}\\

     \includegraphics[width = 1.1\columnwidth]{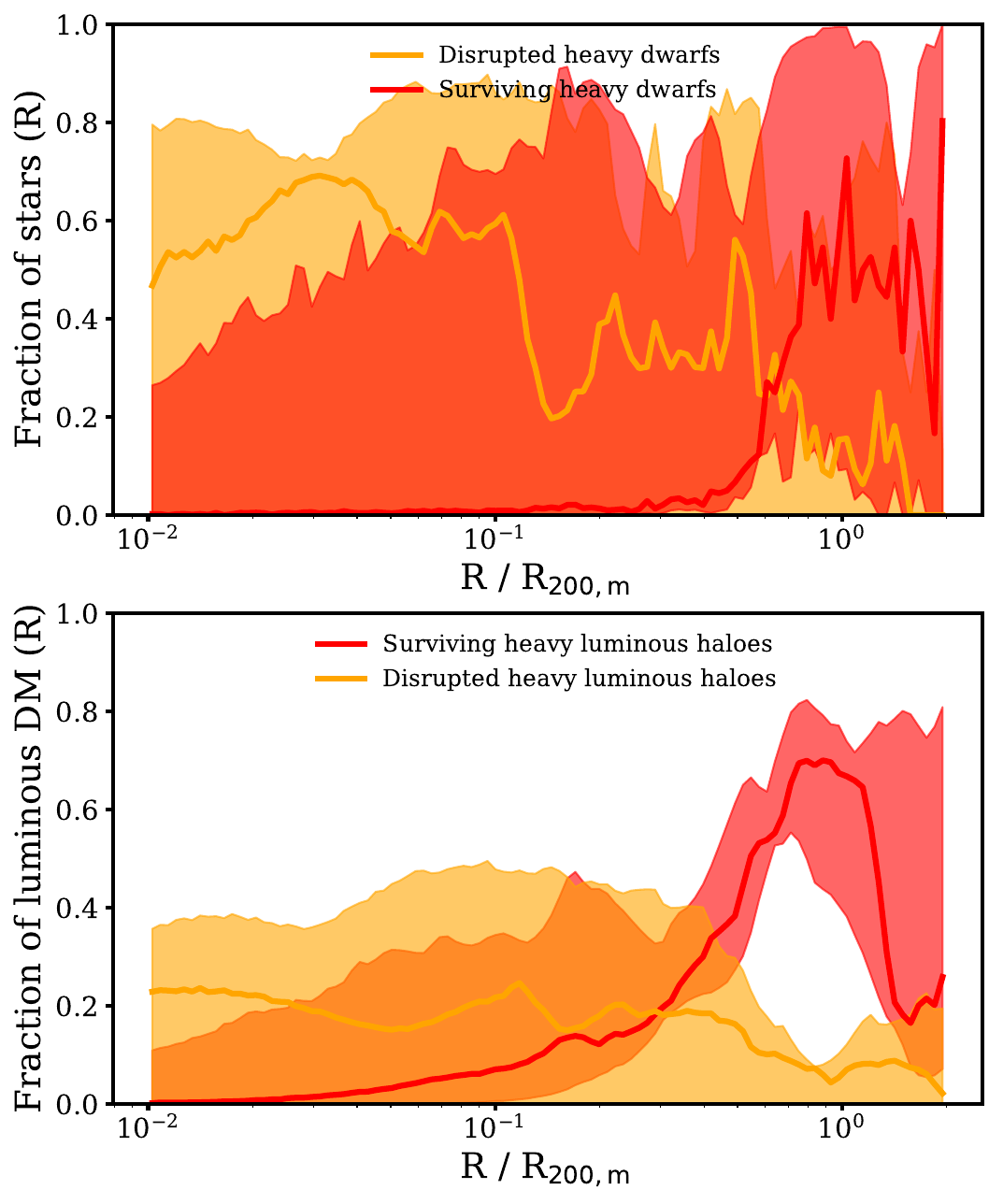}

     \includegraphics[width = 1.1\columnwidth]{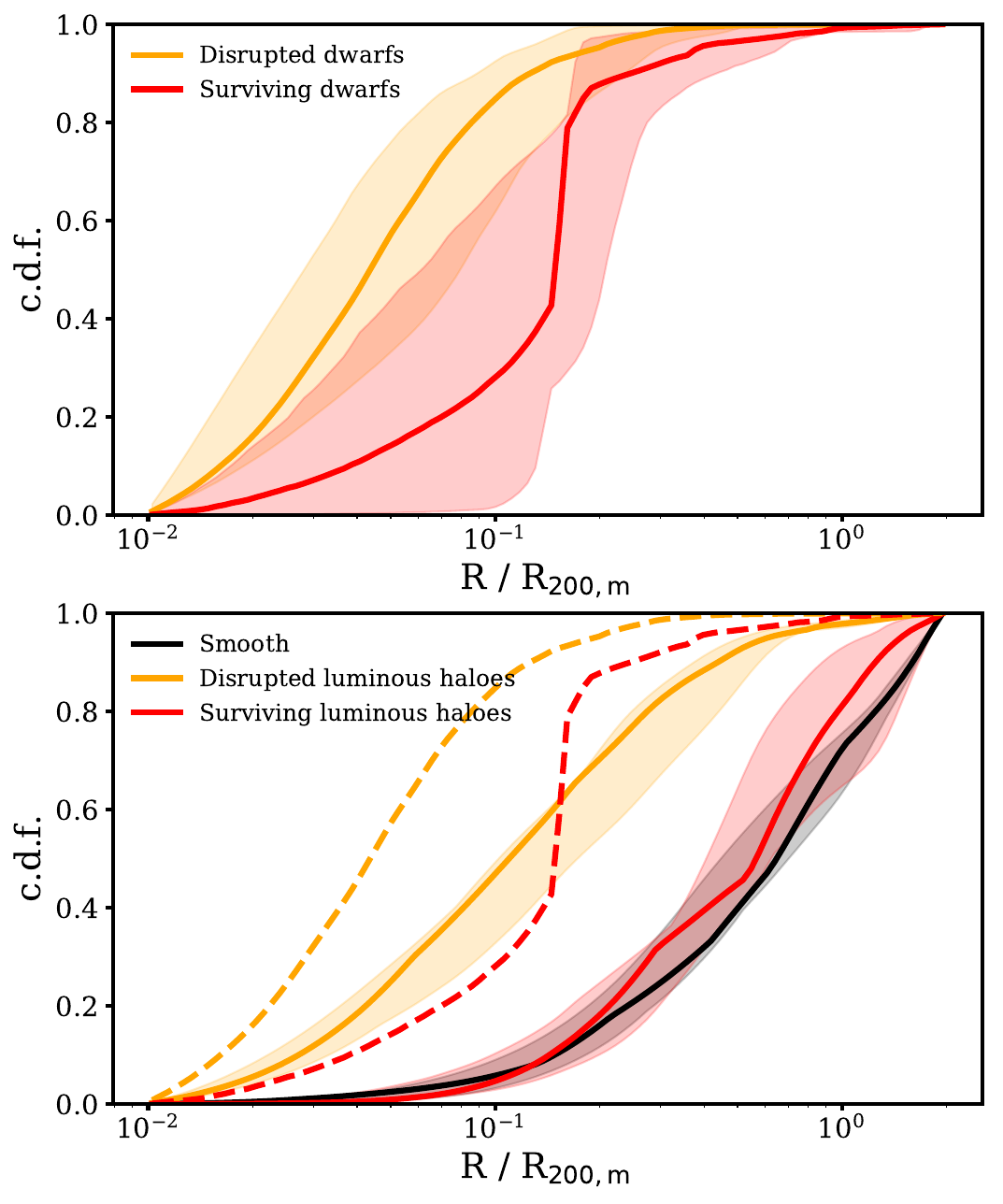}
  
    \end{multicols}
    \caption{Radial distribution of the stellar and dark matter haloes of
      \textbf{\textit{active}} galaxies. The upper panels of each subplot show the
      stars and the bottom panels the dark matter. Yellow identifies
      luminous haloes that are fully disrupted by $z=0$, red shows
      luminous haloes that survive, blue shows disrupted
      dark haloes that never hosted stars, purple shows surviving
      dark haloes that never hosted stars and black shows dark matter
      that was never bound to a subhalo (i.e. the ``smooth''
      component). \textit{Top left:} bar chart showing the fraction of
      the halo made up by each component. The height corresponds to
      the median across the sample of active galaxies and the error
      bars bracket the minimum and maximum values in the
      sample. \textit{Top centre:} The fraction of the halo at each
      radius contributed by each of the components. Solid lines show
      the median of the sample and the band shows ranges for  the entire
      sample. \textit{Top right:} The fraction of the halo at each
      radius made up by disrupted (yellow) and surviving (red)
      \textit{massive} luminous haloes, defined as the dwarfs that make 
      up 50~per~cent of the stellar halo
      ($\sim M_{\rm peak,*} > 10^9$~M$_{\odot}$). \textit{Bottom
        left:} The lookback time at which various halo components as a
      function of radius first entered $R_{200,m}(z)$. \textit{Bottom
        center:} The lookback time at which various halo components as
      a function of radius were stripped from their haloes. For the
      smooth component we display the accretion time as in the panel
      on the left. \textit{Bottom right:} the cumulative distribution
      function of the stars and the dark matter stripped from 
      fully disrupted and surviving dwarfs. In the bottom panel, the
      dashed lines show the median of the stars. }
    
    \label{fig3}
\end{figure*}

\begin{figure*}
    \centering
    \begin{multicols}{3}
    \includegraphics[width = 1.1\columnwidth]{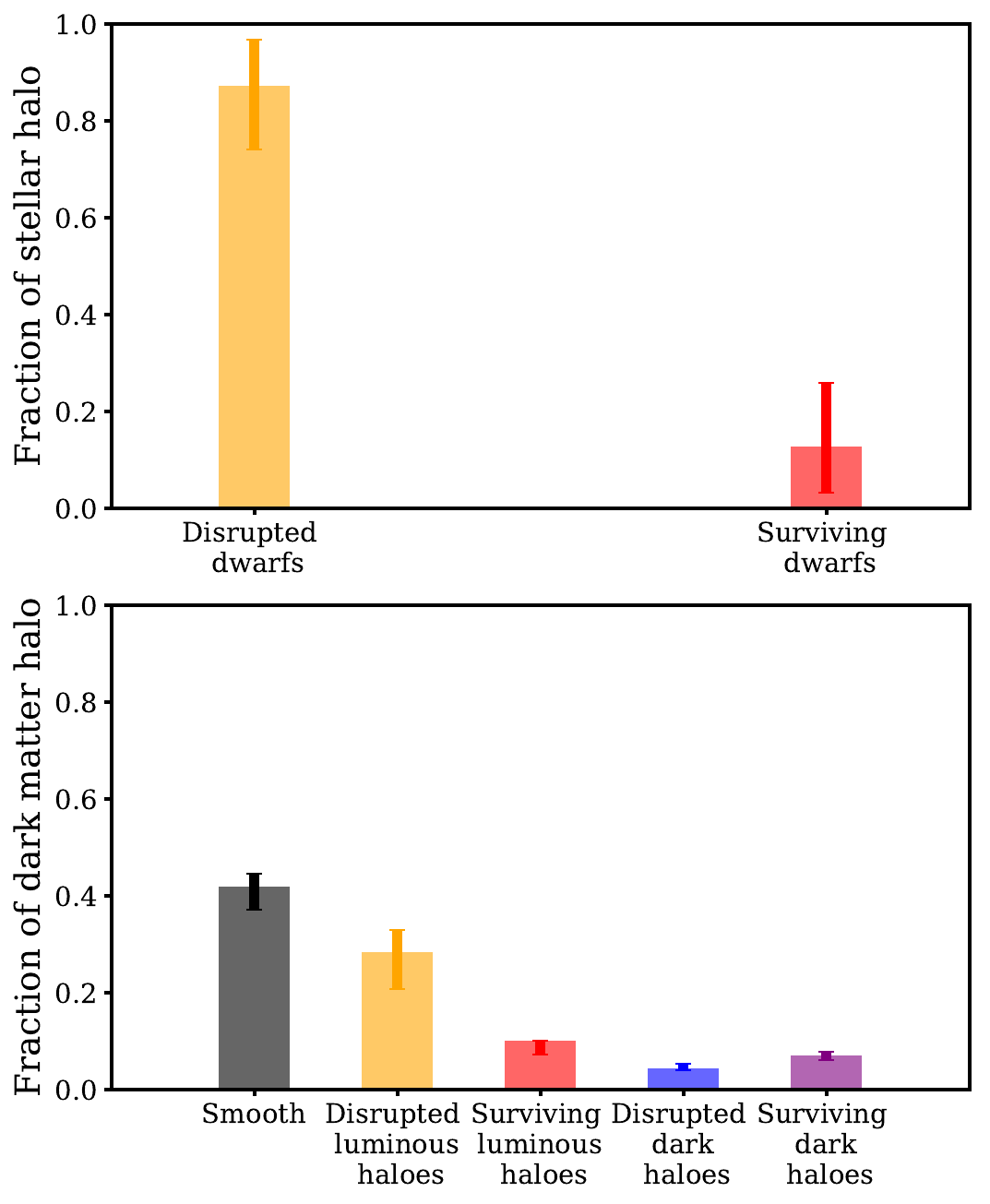}

    \includegraphics[width = 1.1\columnwidth]{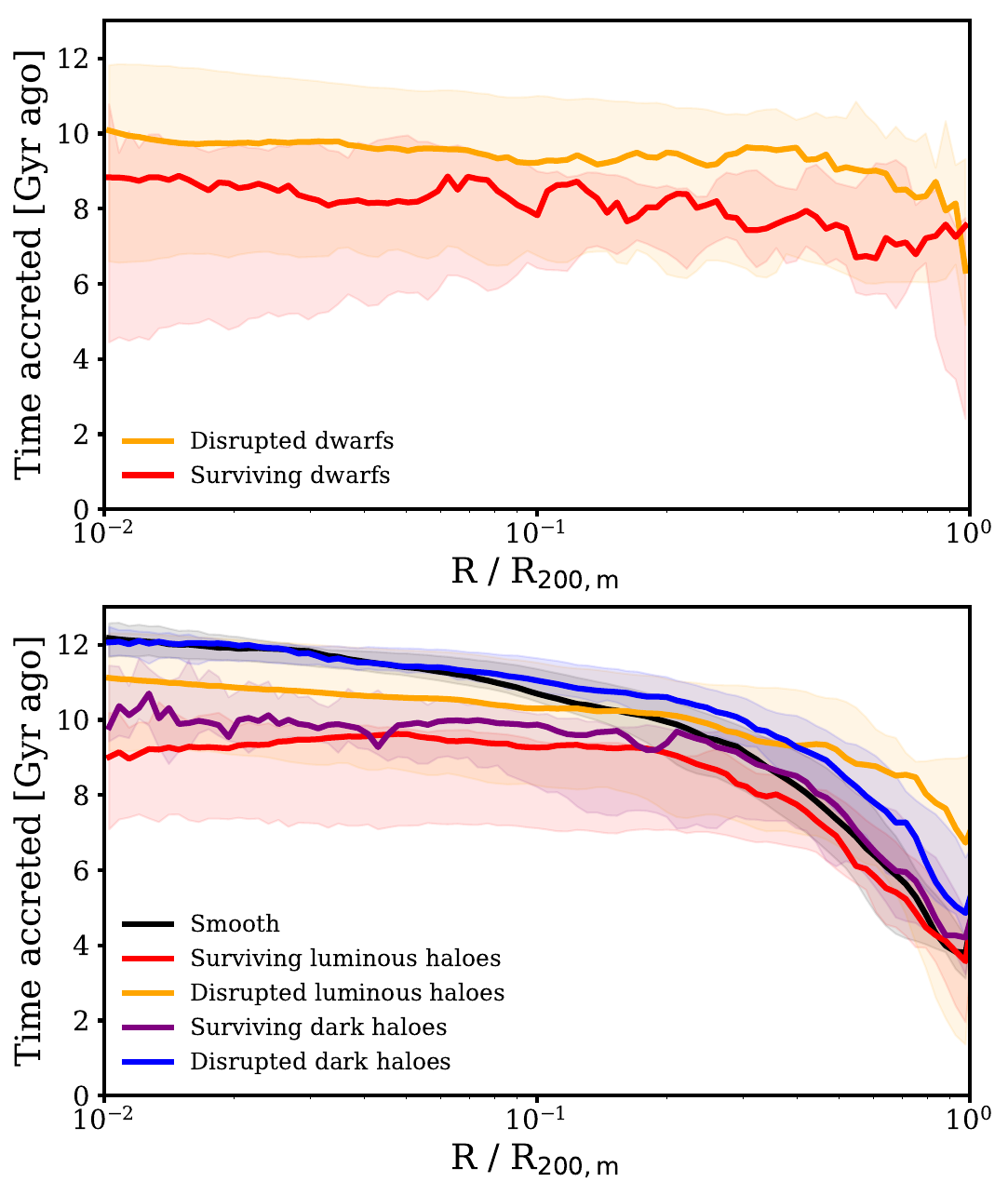}\\

     \includegraphics[width = 1.1\columnwidth]{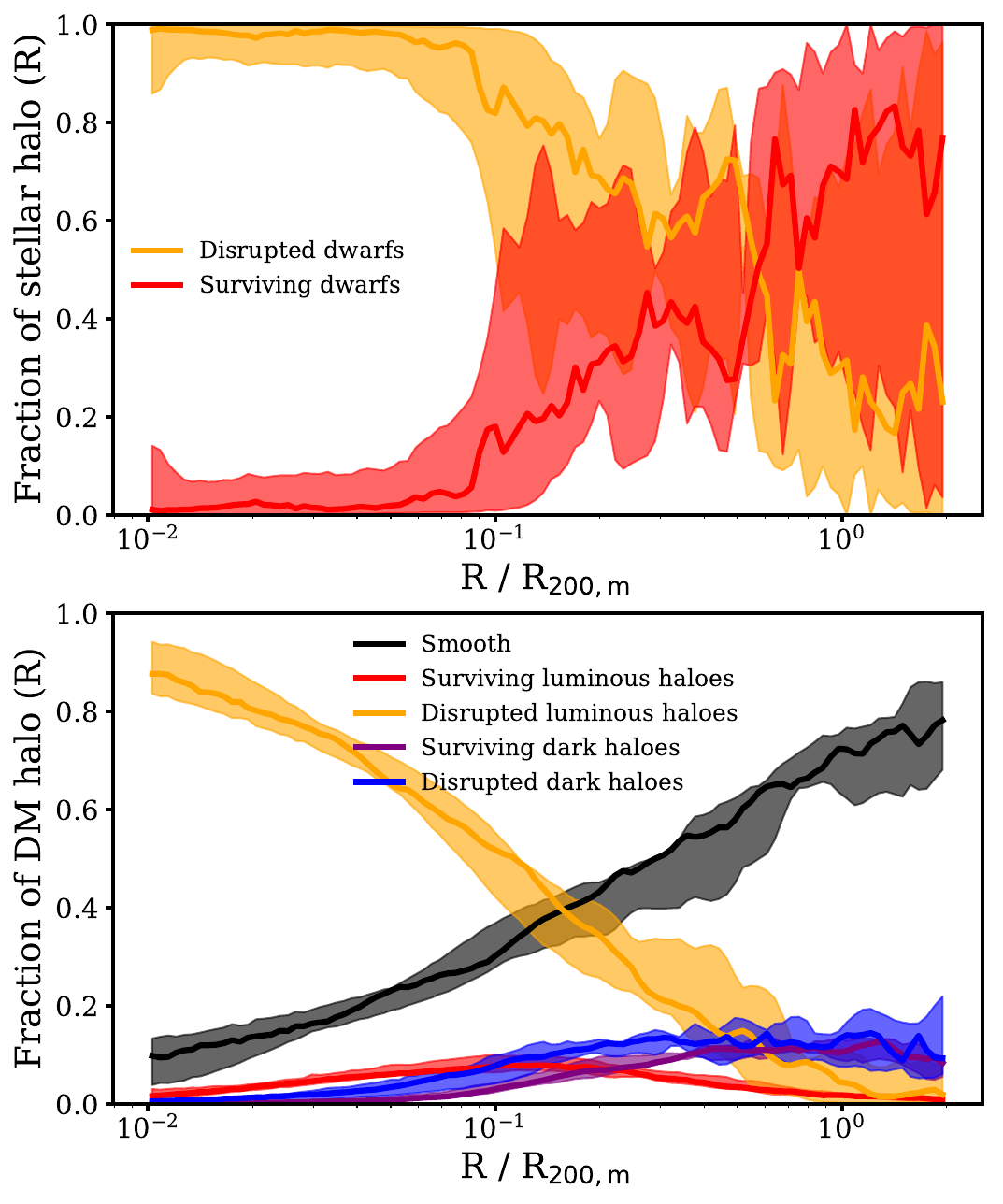}

    \includegraphics[width = 1.1\columnwidth]{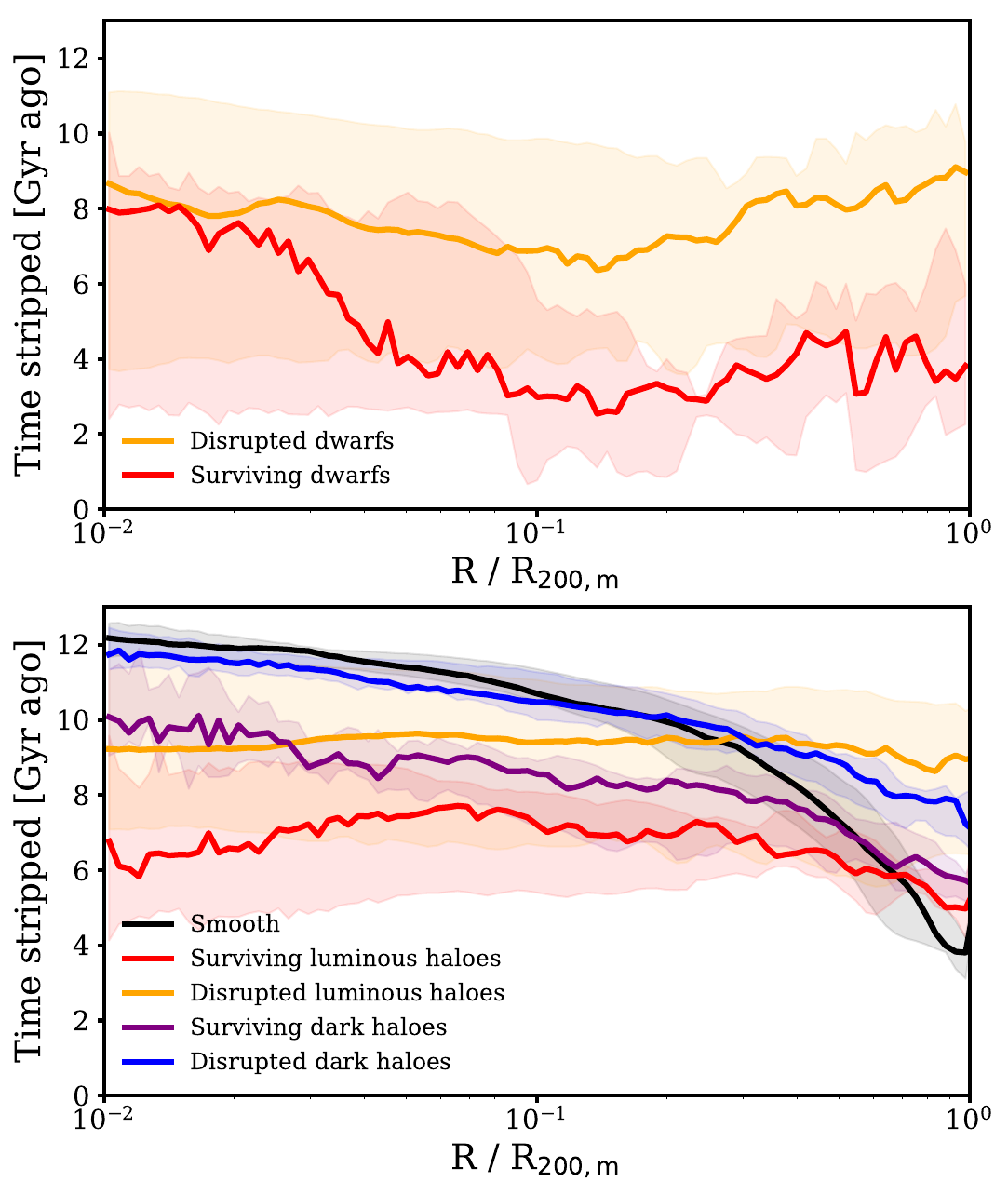}\\

     \includegraphics[width = 1.1\columnwidth]{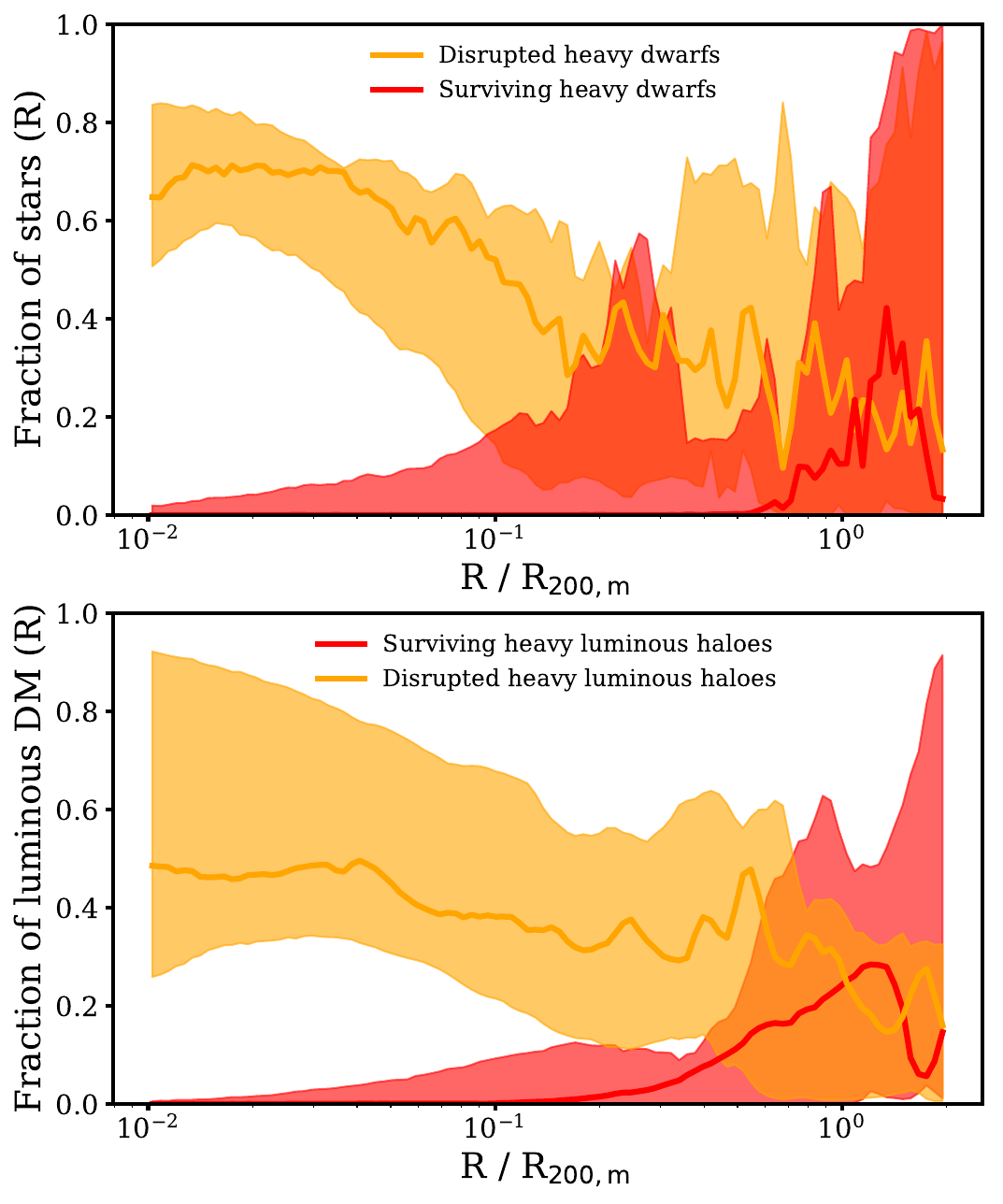}

     \includegraphics[width = 1.1\columnwidth]{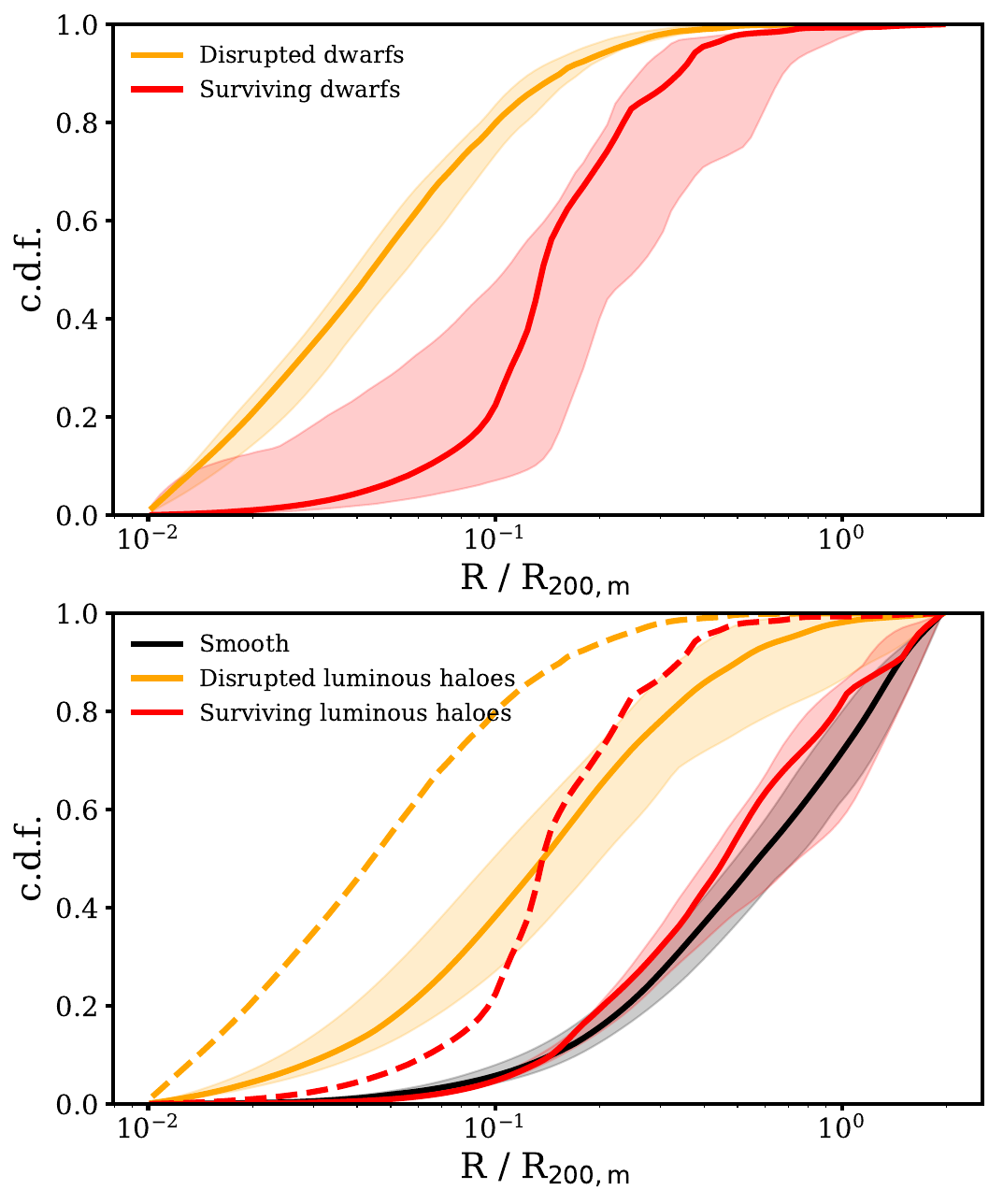}
  
    \end{multicols}
    \caption{Same as Fig.~\ref{fig3}, but for \textbf{\textit{quiet}} Milky Way/M31 analogues.}
    
    \label{fig4}
\end{figure*}

\begin{figure*}
    \centering
    \begin{multicols}{2}
    \includegraphics[width=\columnwidth]{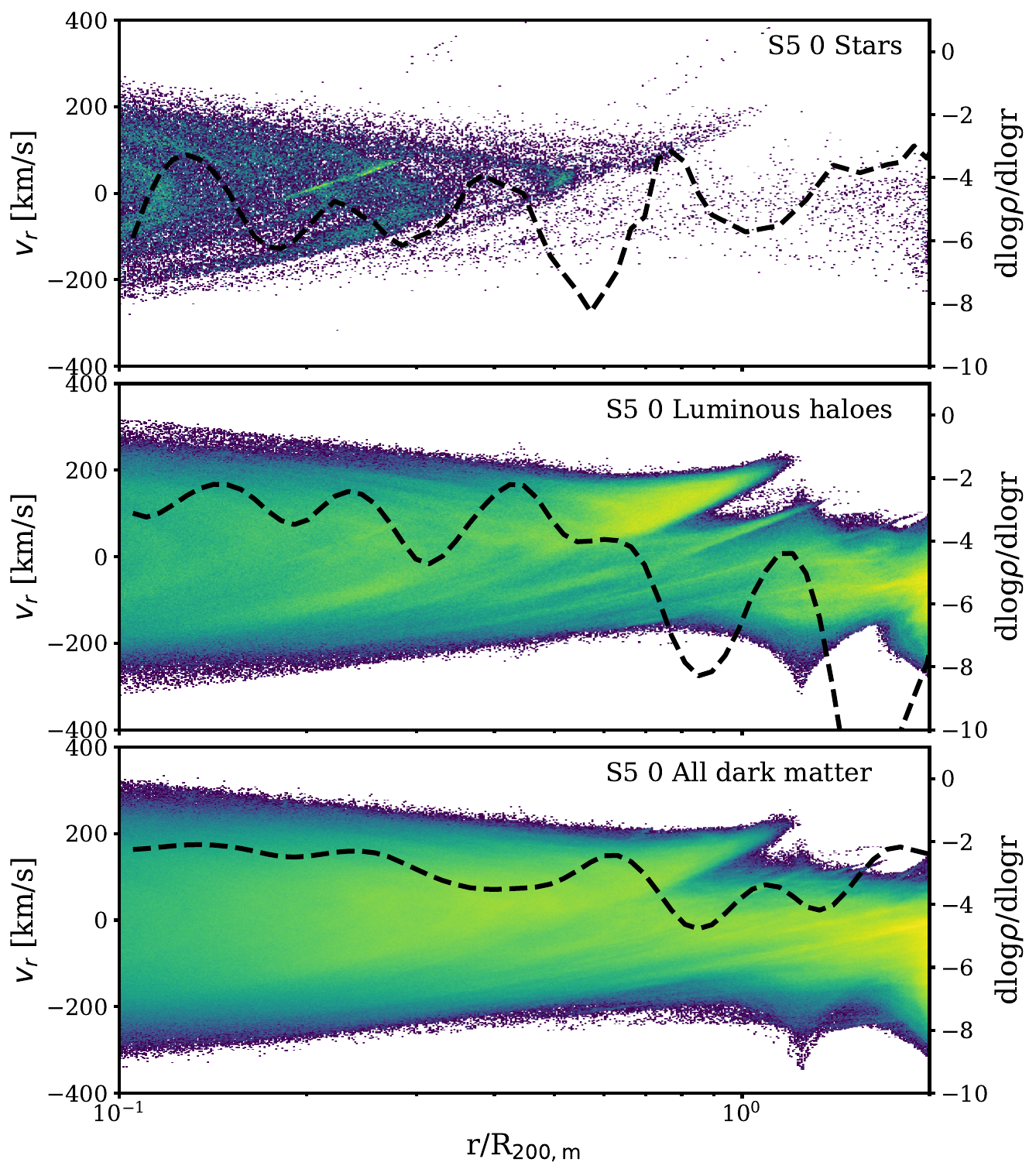} \\
    \includegraphics[width=\columnwidth]{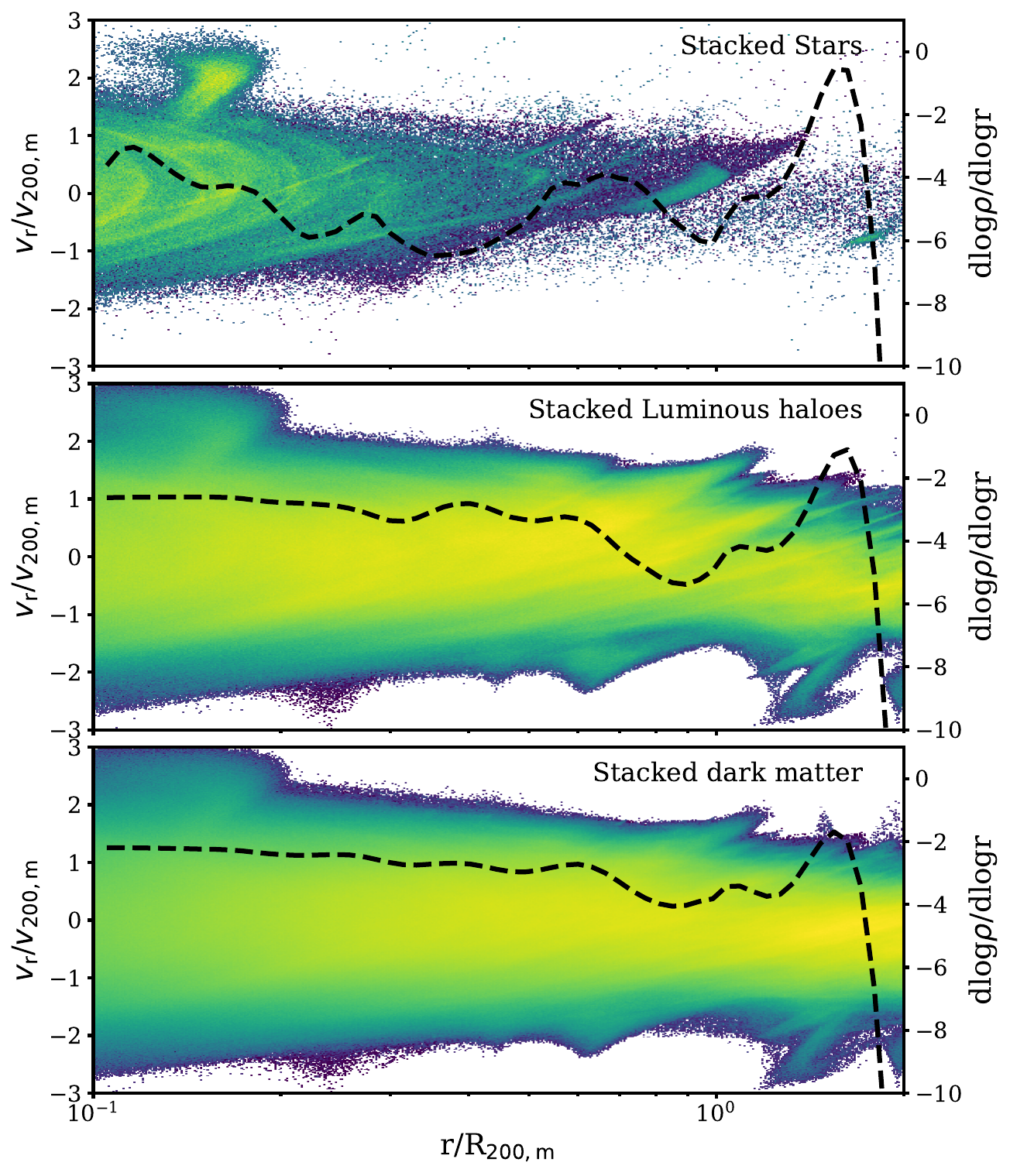}
    \end{multicols}
    \caption{The radial velocity - distance (or ``phase-space'') diagrams of \textbf{\textit{active}} Milky Way/M31 analogues. \textit{Left panel:} individual example of an active galaxy, where in the top we display the entire stellar component, in the middle the dark matter stripped from objects which have in the past hosted stars and in the bottom all of the dark matter in the halo, excluding bound subhaloes. The dashed lines in each case show the log-slope of the density profile. \textit{Right panel:} the sample of active galaxies stacked together. The dashed lines show the log-slope of the density profile of the stack.  The radial velocity axis is normalized by the circular velocity at R$_{\rm 200,m}$.}
    \label{fig5}
\end{figure*}

\begin{figure*}
    \centering
    \begin{multicols}{2}
    \includegraphics[width=\columnwidth]{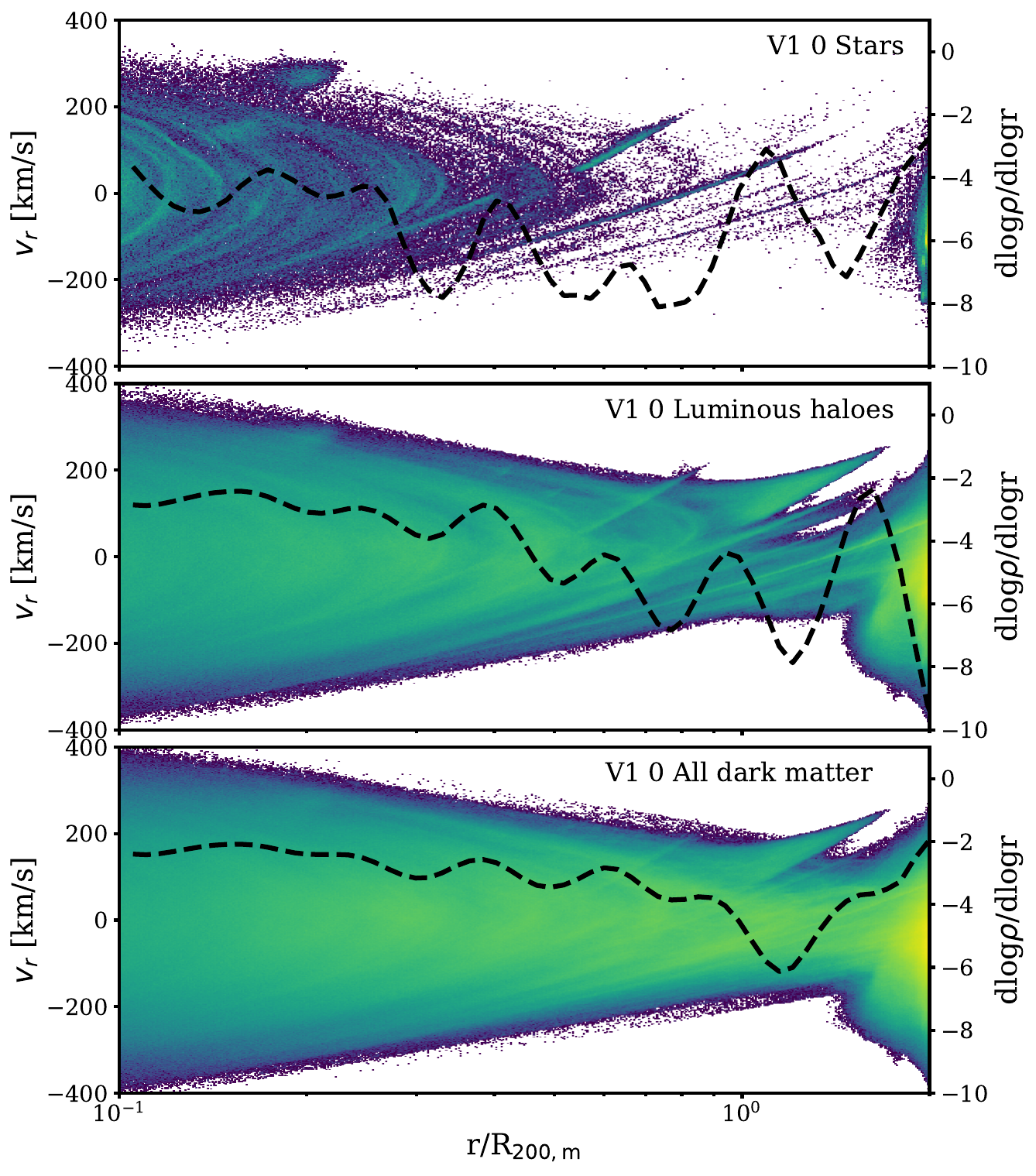} \\
    \includegraphics[width=\columnwidth]{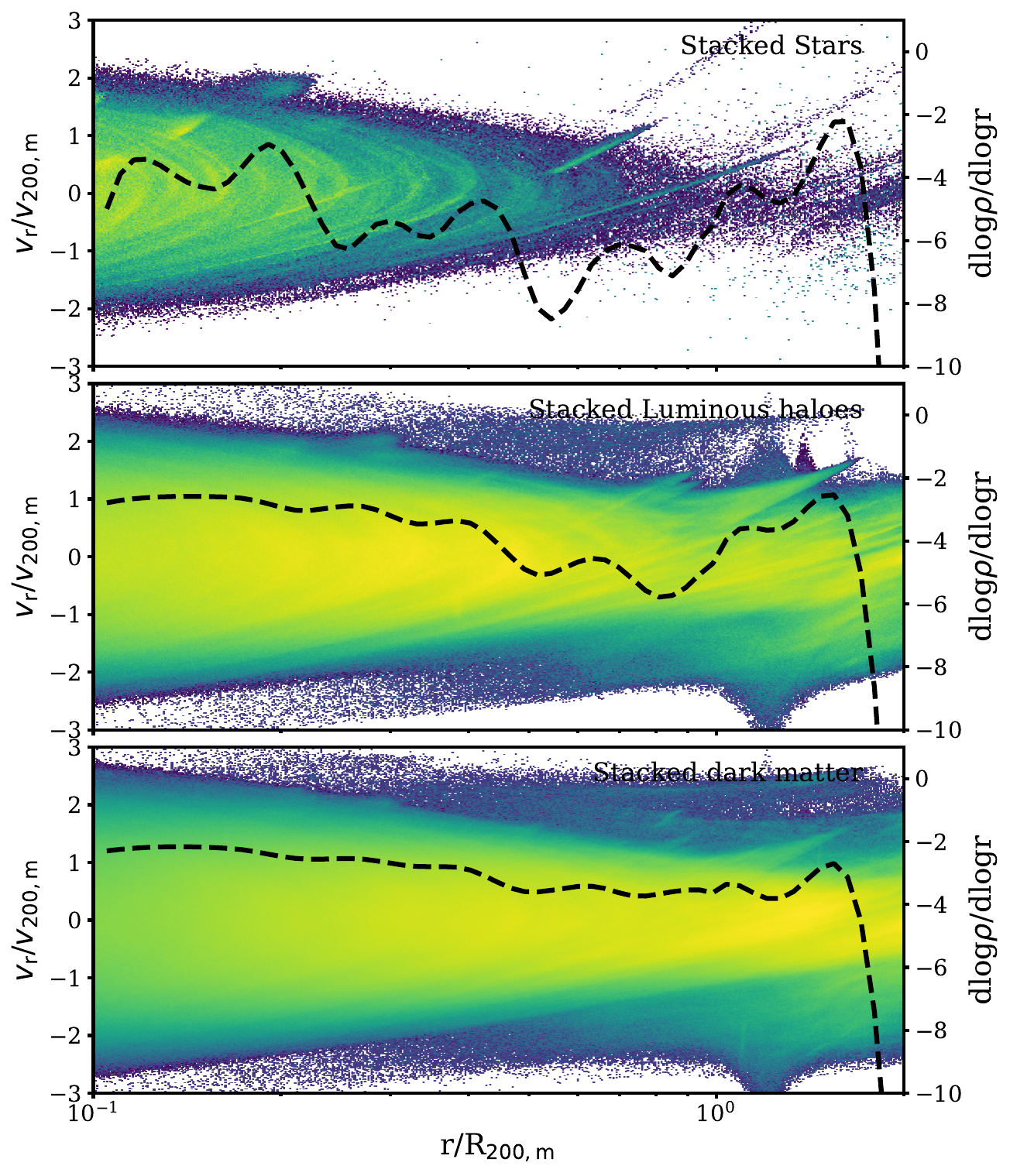}
    \end{multicols}
    \caption{Same as Fig.~\ref{fig5}, but for the \textbf{\textit{quiet}} sample of Milky Way/M31 analogues.}
    \label{fig6}
\end{figure*}

\subsection{Halo composition}

We begin by studying the build-up of the accreted stellar and dark matter haloes of Milky Way-mass galaxies. Dark-matter-only (DMO) simulations show that major mergers (mass ratio $> 1:10$) contribute only 20 per~cent of the dark matter halo mass in Milky Way-mass galaxies, with the majority made up of the `diffuse' component, which includes unresolved haloes and minor mergers in roughly equal amounts
\citep{wang_aq}. However, this picture need not be reflected in the
stellar halo because only the most massive haloes will form a galaxy
which can subsequently be deposited into the stellar halo of the Milky
Way. The power-law form of the CDM mass function implies that
small dwarfs are more abundant than large ones; however it is the less
numerous, more massive, haloes that contain the most stellar
mass. This opens up the question of whether most of the stellar halo
is made up of stars gained in mergers with many small dwarfs or few
large ones and whether these are surviving or disrupted at the present day. These questions have in the past been tackled with semi-analytical approaches \citep{cooper,deason_mao_wechsler, delucia_helmi}, as well as hydrodynamics simulations \citep{monachesi_2019, fattahi_two_pops, fire_halo}. These have generally favoured Milky Way stellar haloes that are dominated by stars from a few, massive dwarfs that are already disrupted. In the following, we investigate whether these findings hold in the APOSTLE simulations. We emphasize that in this work we only consider the the stellar halo stars contributed by the accreted dwarf galaxies and ignore the \textit{in-situ} component contributed by the stars scattered from the Milky Way disc \citep{belokurov_2020}. A number of previous works have found that this component becomes subdominant beyond $\sim 50$~kpc \citep{zolotov_halo_insitu, font_insitu, artemis_halo,fire_halo}, we therefore exclude it from our analysis, which instead focuses on understanding the origin of the phase-space features defining the "edge" of the stellar halo.  

\subsubsection{Stellar halo composition}

In Fig.~\ref{fig1} we show the make-up of the stellar and dark matter
haloes in terms of the peak (that is the maximum value attained)
stellar and dark matter masses of infalling dwarfs (upper left) and in
terms of the number of dwarfs at a given peak mass (upper right). The
peak in the fraction of the stellar halo contributed by objects of a
given mass occurs at peak stellar mass of
$M_{\rm peak, *}\sim 10^9$M$_{\odot}$, where, on average, dwarfs with
this stellar mass make up $\sim 40$~per~cent of the stellar
halo. However, it can also be seen that the scatter is rather large
and some of the Milky Way analogues either accreted no objects of this
stellar mass, or only accreted them at late times, when the amount of
stellar stripping was insufficient to contribute any significant mass
to the halo (surviving dwarfs are shown by the dotted red line). On the
bottom left of Fig.~\ref{fig1}, we show the cumulative version of this
figure. Evidently, dwarfs with peak stellar mass of above
$10^8$M$_{\odot}$ contribute at least 70~per~cent of the stellar
halo. We also note that typically less than 20~per~cent comes  from
dwarfs surviving at $z=0$. The stellar mass function of the dwarf
progenitors (bottom right of Fig.~\ref{fig1}) suggests there are
approximately 20 progenitors with stellar mass above $10^8$M$_{\odot}$
(7 surviving) and 6 above $10^9$M$_{\odot}$ (3 surviving) in a
typical Milky Way analogue. These results are consistent with the work of \citet{elias_2018} that used Illustris simulations, but there are also differences in two aspects. One is that they find no contributing satellites above $\sim10^{9.5}$~M$_{\odot}$ in stellar mass, while we have 1-2 contributors. We attribute this difference to their sample selection of Milky Way-mass galaxies that lack a massive LMC-like satellites at the present day. Another difference is that below $\sim 10^7$~M$_{\odot}$ their contributing number of satellites is higher than ours. We believe that this is due to our choice of not counting dwarf galaxies that merge with a bigger galaxy within the Milky Way. \citet{elias_2018} also have generally lower counts of surviving satellites at all masses than presented in this work, which we believe is a consequence of mass resolution in their simulations, which is a factor of $10^2$ lower and likely results in enhanced satellite disruption \citep{vandenbosch2}. Conversely, the work of \citet{fattahi_two_pops} using AURIGA simulations returns mass functions that have a similar shape to ours, but have a systematically lower normalization for both surviving dwarfs and all progenitors, while still agreeing with our results within the scatter of Milky Way analogue mass functions. Their method of counting dwarfs contributing to the halo is very similar to ours, so the discrepancy likely arises from the fact that these authors only count the dwarfs accreted after $z=3$ \footnote{This choice stems from unreliability of determining the main progenitor at early times. In this work, we rely on the definition of the main progenitor by the \textsc{HBT+} halo finder, which is based on the mass and specific kinetic energy of progenitor candidates.} and that the discs in AURIGA are more effective in disrupting the satellite population than APOSTLE \citep{richings}, though the differences are important for substructures with $M_{\rm DM} < 10^8$~M$_{\odot}$ which should in principle remain dark. In any case, we have verified that restricting our count of halo contributors to those infalling after $z=3$ places our results in agreement with \citet{fattahi_two_pops}.

\subsubsection{Dark matter halo composition}

We now focus on the dark matter. Immediately, we see that the tallest
peak in dark matter halo contribution is at $\sim 10^4$~M$_{\odot}$,
which is the dark matter particle resolution in our simulations. This
is the contribution of the `smooth' and `unresolved' component. Among
the resolved bound substructures, it can be seen that significant
fractions of the dark matter halo mass come from dwarfs with peak halo
mass $\sim 3\times10^{10}M_{\odot}$. The purple histogram highlights
subhaloes that have hosted stars in the past. It is evident that all
objects above $\sim 10^{10}M_{\odot}$ hosted stars in the past. The
cumulative version of this plot (bottom left of Fig.~\ref{fig1})
confirms the previous findings that nearly half of the dark matter
halo is in a `smooth' component and the other half in bound
structures. It can also be seen that about 40~per~cent of the dark
matter halo mass has come from objects that have hosted stars in the
past and of those, nearly all had halo mass  above
$10^{9}M_{\odot}$, consistent with the hydrogen cooling limit
arguments, while 30~per~cent of the halo mass comes in subhaloes of 
peak mass greater than $10^{10}M_{\odot}$. The dark matter haloes of
Milky Way-mass galaxies had 100-200 luminous contributors (with
stellar masses $\gtrsim 10^3 M_{\odot}$), though only 10-20 of them
make up the majority of the stellar halo.

\subsection{Active and quiet dark matter halo assembly}

Before we proceed to examine the radial distributions of various
components of the stellar and dark matter haloes, we split our sample
of Milky Way and M31 analogues into those with `active' formation
histories (haloes which are still rapidly increasing in mass down to
$z=0$) and those with `quiet' formation histories (those whose halo
growth rate over the past few gigayears is slow). We make this sample
separation for two reasons. One is that we expect that the radial
distributions of debris from surviving and disrupted dwarfs will be
different in these two cases -- the locations of stars coming from
surviving dwarfs are expected to peak away from the centre of the
galaxy and the disk. Secondly, the sample separation is motivated by
the distinction in the inferred formation histories of the Milky Way
and M31. The Milky Way is believed to have been relatively quiet
\citep{deason_mw_m31, pillepich, lachlan}, while M31 is still actively
assembling. In Fig.~\ref{fig2} we show the dark matter halo assembly
histories of our Milky Way/M31 analogues; the subsamples are shown in
blue (active) and red (quiet).  The quiet sample is characterised by
analogues that had formed $\sim80$~per~cent of their mass 6~Gyr ago
and had a slow growth rate after that. In contrast, the `active'
halos formed $\sim80$~per~cent of their mass approximately 3~Gyr ago
and their growth thereafter is fast. We find that lower-mass haloes in our sample tend to be `active', while higher-mass haloes are typically `quiet'. This may seem surprising from the hierarchical structure formation considerations. This occurs because our ``zoom'' simulations are constrained to have haloes in the Milky Way / M31 mass range. If we instead define recent accretion history as that within the last 8.5~Gyr \citep{diemer_and_kravtsov}, we recover the expected trend where high-mass haloes have more active assembly histories. We find no correlation between the assembly histories of the two main haloes in each simulation volume.

There are two extreme examples in both of these categories. Within the
`active' sample there is a galaxy that has built up nearly 50~per~cent
of its halo mass in the last 5~Gyr. This is due to a very recent major
merger. This galaxy is an outlier in this category, but is, of course,
formally `active'. Within the `quiet' sample we see a galaxy which had
two large mergers $\sim7$~Gyr ago and has since assembled only
about ~10~per~cent of its final mass. This is distinct from the rest
of the `quiet' sample, where nearly all of the final halo mass has
been built up $\sim9$~Gyr ago. In the following, we keep these
galaxies in their respective categories, bearing in mind that the rest of
galaxies in each sample have very similar assembly
histories, so that our outliers will likely not affect the median
radial distributions, but would instead contribute to the scatter.

\subsection{Radial distribution of halo components}

We have so far shown that the stellar halo of Milky Way-like galaxies
is dominated by stars stripped from a few massive dwarfs that are
primarily disrupted by $z=0$, while the dark matter halo is dominated
by a smooth, unresolved component. We summarize these findings in
the top left of Figs.~\ref{fig3} and \ref{fig4} for the active and
quiet Milky Way~/~M31 analogues, respectively. It is clear that the
smooth component is the dominant contribution to the dark halo. In
both active and quiet samples the disrupted dwarfs dominate the
stellar halo, but surviving dwarfs contribute more significantly in
the active sample. Interestingly, while the mean contributions to the
stellar halo of the disrupted and surviving dwarfs are similar in the
quiet and active samples, there are clear differences in the contributions 
of these objects to the dark matter. Specifically, disrupted and surviving
dwarfs contribute roughly the same amount of dark matter in the
`active' sample, yet they contribute very different fractions to the
stellar halo. This suggests that substantial amounts of dark matter
have been stripped from surviving dwarfs, but their stellar component
has not been significantly affected. We now investigate how these
components are distributed within the haloes. We aim to establish, for
instance, whether the smooth halo component is dominant at all radii
or only in the outskirts of haloes and whether these radial distributions
are different in active and quiet Milky Way~/~M31 analogues.

In the top-centre panels of Figs.~\ref{fig3} and \ref{fig4}, we show
the fraction of the stellar and dark matter haloes that each component
contributes at a given radius. In both active and quiet samples,
disrupted dwarfs make up almost all of the stellar halo up to
0.1$R_{200, \rm m}$ and remain dominant out to $\sim 0.8 R_{200,m}$ with scatter, while surviving dwarfs contribute most of the material outside that radius. This is an agreement with fig.6 of \citet{fattahi_two_pops}, who find that massive, destroyed dwarfs dominate the contribution out to $\sim$100-200~kpc in AURIGA Milky Way analogues ($\sim 0.3-0.5 R_{200,m}$ in our simulations).

In the active sample, there is a small radial range between
0.1 and 0.2~$R_{200, \rm m}$ where material stripped from surviving dwarfs
dominates on average, providing $\sim65$~per~cent of the visible
matter in the halo. This is material stripped recently (within the
last 2~Gyr) from surviving dwarfs that wandered close to the centre of
the halo. Interestingly, we do not see a corresponding peak in the
dark matter contributed by surviving luminous haloes. The reason for
this is the outside-in stripping of infalling dwarfs (illustrated in
the bottom panels of Figs.~\ref{fig3} and \ref{fig4}), that makes the
stripping times of the dark matter and the stars different.

For the dark matter, in both active and quiet samples, we see that the
smooth component makes up only a small fraction of the inner dark
matter halo ($\sim10$~per~cent at 0.01$R_{200, \rm m}$, roughly the Milky
Way half-light radius); the majority of the dark matter has come from
the dark matter of disrupted dwarf galaxies. The smooth component
becomes dominant at $\sim0.2R_{200, \rm m}$, but the dark matter from
disrupted dwarf galaxies is still a major contributing component out
to $\sim0.3 R_{200, \rm m}$ in the active sample and $\sim0.6 R_{200, \rm m}$ in
the quiet sample.

\subsection{Radial contribution by mass}

We have now established that it is the most massive, luminous dwarfs
that make up the majority of the stellar halo. However, it is still
unclear whether this is true over the entire radial extent of the
halo. It could be the case that the massive dwarfs dominate only in
the centre, where most of the stars are expected to be deposited in a
merger, while the more numerous small dwarfs with longer dynamical
friction sinking timescales deposit their stars in the outer halo,
dominating the local stellar content. To address this question, we
split the disrupted and surviving dwarfs by mass, such that massive
dwarfs above a given threshold mass make up 50~per~cent of the stellar
halo. We then show the radial contribution of each component to the
stellar halo. 

In agreement with the upper left panel of Fig~\ref{fig1}, we find that
this threshold peak stellar mass is typically $\sim10^9$~M$_{\odot}$
and ranges between $10^{8}-10^{9.5}$~M$_{\odot}$ for our Milky
Way~/~M31 analogues, corresponding to LMC/SMC-mass dwarf galaxies. It
can be seen that massive dwarfs, surviving and disrupted, contribute
over $\sim30$~per~cent of the mass at all radii. The massive disrupted
dwarfs clearly dominate the stellar halo content within
$0.1R_{200, \rm m}$. The contribution of the most massive dwarfs diminishes
between 0.1 and 1$R_{200, \rm m}$. Indeed, the smaller dwarfs become more
important in the `intermediate' halo regions. This is likely because
their orbits are relatively more tangential due to the reduced effects
of dynamical friction. We do find that the debris stripped from lower-mass dwarfs has {\it on average} larger orbital apocentres, however the most massive haloes typically have larger {\it maximum} apocentres due to their larger size and velocity dispersion. It is also clear that the peak in the
contribution of surviving dwarfs seen between 0.1 and 0.2 R$_{200,\rm m}$
in active galaxies (top-centre panel) is not caused by the most
massive dwarfs that entered the halo. This is not surprising, given
that in Fig.~\ref{fig1} we saw that the peak masses of disrupted
dwarfs are on average greater than those of surviving ones. In other
words, it is likely that the most massive contributor to the Milky
Way's stellar halo (in terms of peak stellar mass) has already been
disrupted.

This picture is somewhat different in the dark matter, where the most
massive mergers contribute typically no more than $30-40$~per~cent of
the dark matter coming from luminous subhaloes. This is most likely a
consequence of the stellar mass--halo mass relation \citep{behroozi, moster}, whereby dark matter dominates the mass of smaller
dwarfs (e.g. dwarf spheroidals and the ultra-faints) more so than in the
dwarf irregulars (e.g LMC). As a result, smaller dwarf galaxies contribute a larger fraction of their mass in dark matter than the larger dwarfs. Overall, it is remarkable that out to the outermost radius within the halo, the material of 3-4 massive dwarfs can make up nearly half of the stellar halo.

\subsection{Radial gradients in accretion and stripping time}

We now examine the radial accretion time gradients in the stellar and
dark matter haloes, both regarding the time of entry into 
$R_{200,m}$ and the time when the material was stripped from infalling
subhaloes\footnote{ The stripping time is not defined for
  the smooth halo component, so instead we use the time when these
  particles entered $R_{200, \rm m}(z)$.} (lower panels in Figs~\ref{fig3} and~\ref{fig4}). 

The radial distribution of the stellar accretion times shows a
remarkably flat profile both for disrupted and surviving dwarfs out to
large radii, in both subsamples. This either suggests that the
majority of dwarfs contributing to the stellar halo were accreted at
roughly the same time or that the stellar halo is dominated by one or
two objects. In the quiet sample, this may be the case, as few
disrupted dwarfs make up nearly half of the stellar halo at all radii. 

For the active sample, above $\sim 0.4-0.5R_{200, \rm m}$ the stellar halo
has an increasing contribution from more recently accreted objects,
both surviving and disrupted. This is in agreement with the findings of \citet{font_insitu}, using GIMIC simulations, who find a declining profile in accretion time as a function of radius for their Mily Way analogue. When looking at the radial stripping
time (bottom centre of Fig.~\ref{fig3}), we begin to see a mild
gradient, whereby material near the halo centre was stripped earlier
and the material in the outskirts later on, before the slope flattens,
such that above 0.2$R_{200, \rm m}$ the halo material was stripped at
approximately the same time. This suggests either that the material
deposited in the outer parts of the halo from a given dominant object
was stripped later on (which seems unlikely as we expect the infalling
dwarfs to spiral towards the centre and deposit their stars there), or
that the outer regions of the halo are dominated by dwarf galaxies
that came on wider orbits, allowing their stars to be stripped later
and at larger radii.

Overall, for the disrupted dwarfs we observe an offset between the
accretion and stripping times of approximately 2~Gyr, while for
surviving dwarfs this difference amounts to $\sim5$~Gyr, suggesting that
the surviving dwarfs are harder to strip due to the particulars of
their mass, the mass of the Milky Way analogue and the orbit.

For the dark matter, we see that the `smooth' component in the inner
regions is likely of primordial origin, while in the outer regions,
where it dominates, the smooth component was deposited into the halo
more recently. The latter is, however, also true for other
constituents of the dark matter halo, where the outer regions are made
up of dark matter accreted recently. Interestingly, the radial
distributions suggest that in the outermost regions the dark matter
particles had been stripped prior to entering $R_{200, \rm m}$. We also see
this odd behaviour for the stars that came from dwarfs that are
disrupted. 

This may be in line with the findings of \cite{wang_aq}, who suggest
that some of the "smooth" dark matter component in the halo could have
come from previously bound structures that lost dark matter particles
during a merger prior to infall. Moreover, considering that we are
analysing Local Group-like environments, it is plausible that some of
the material could have come from subhaloes that had been previously
stripped within the companion halo, then ejected, before infalling
into the Milky Way analogue. However, we have found that these objects
make up no more than 10~per~cent of the material currently outside
$R_{200,\rm m}$. 

Instead, there are two main sources for the apparent inconsistency
between the `stripping' and the `accretion' time. The first
contributes at small and large radii and is due to dwarfs that merge
with the Milky Way analogue at early times, when R$_{200,\rm m}(z)$ is
rather small and thus it takes a while for some of the dark matter
particles, which become unbound during the merger, to formally cross
R$_{200, \rm m}(z)$ and therefore to be identified as having been
``accreted''. This is particularly likely when the incoming dwarf
itself is in the process of assembly and has an extended halo of
loosely bound material. The second source is more important in the
outermost regions of the halo and consists of massive, LMC-like,
dwarfs that have entered the halo recently, so that some of the
particles still have not crossed R$_{200,\rm m}(z)$. In both cases, the
wide spread in particle binding energies in these massive dwarfs can
lead to a sufficiently small tidal radius, and the dark matter can
become unbound prior to crossing R$_{200, \rm m}(z)$.

It is additionally interesting to note that the discrepancy between
the accretion and stripping times is more pronounced in active
galaxies than in the quiet sample. This may explain why the
disrupted dark haloes contribute more in the active galaxies near
$R_{200, \rm m}$ than they do in the quiet sample. As the active sample
grows significantly in dark matter in the last 3~Gyr, it is likely
that these dark haloes were `pre-processed' by large objects that
entered $R_{200, \rm m}$ at late times.

It is clear that the central component is made up of stars stripped
earlier, while the outer halo is made up of stars stripped later,
which is also the case for the dark matter. However, one can also see
that all radii, for disrupted and surviving dwarfs, the dark matter is
stripped earlier than the stars. In the bottom right of
Figs.~\ref{fig2} and \ref{fig3}, we compare the cumulative
distributions of stars and dark matter stripped from disrupted and
surviving dwarf galaxies. It can be seen that the stars are
significantly more centrally concentrated than the dark matter, with
a half-mass radius of $\sim0.05 R_{200,\rm m}$ for the disrupted dwarfs,
compared to $\sim0.1 R_{200,\rm m}$ for the dark matter. Similarly, half
of the mass in stars stripped from surviving objects is at
$\sim0.15 R_{200,\rm m}$ for the stars and $\sim0.5 R_{200,\rm m}$ for the
dark matter. This reinforces the idea that dark matter in dwarf
galaxies typically gets stripped earlier and more efficiently than the
stars.

\section{Phase-space features in the stellar and dark matter haloes}
\label{phasespace}

\subsection{The radial velocity -- distance diagram}

In this Section we examine some of the orbital properties of the stars
and the dark matter in the halo. On the left of Fig.~\ref{fig5} we
focus on one example of an active halo. In the top panel, we show the
$v_r-D$ diagram for all the stars in the halo (i.e. stars stripped
from dwarf galaxies). In the centre we show dark matter that has come
from luminous haloes and at the bottom we show all dark matter,
including the smooth component. Similarly, in the left panel of
Fig.~\ref{fig6}, we show an example of a quiet galaxy. The dashed
black lines show the computed log-slope of the density profile. These
log-slope profiles are of interest in establishing the ``splashback''
radius of the dark matter and the ``edge'' of the galaxy,
corresponding to minima in the log-slope of the profiles.

In order to compute the log-slope profiles, we follow the procedure of
\cite{edgeofthegalaxy}. Namely, we bin the particles in 75 radial
$\log_{10}$-spaced bins and in 11 angular-spaced bins. Since we
consider Local Group-like systems, the splashback radii of the two
main haloes may overlap. In order to circumvent this, we discard the
angles $\cos(\theta) <-0.6$ and $\cos(\theta) > 0.6$ measured from a
vector joining the two halo centres. For each radial bin, we then take
the median of all bins in angle. Furthermore, we apply the
fourth-order Savitzky-Golay filter \citep{savitzky} over the 15 nearest
bins to smooth the density profiles and compute the log slope. We use
the same number of bins for the stars and for the dark matter. Note
that we use a greater number of radial and angular bins than
\cite{edgeofthegalaxy}, who analysed a lower-resolution version of
APOSTLE\footnote{Since we take a median of the angles, the `edge' of
  the galaxy we infer is defined by particles that are generally well
  phase-mixed by the present day, rather than by potentially 
  denser but highly anisotropic particle distributions arising from
  very recent accretion events \citep{mansfield}. We have also carried
  out convergence studies, varying the number of radial and angular
  bins and found that the locations of the steepest log-slope
  minima do not vary significantly so as to affect the
  conclusions of this work.}.

Examining the two examples visually, it is clear that `quiet' galaxies
are more structured in phase space, with clear `shells' of particles
moving on similar orbits. The `active' galaxy exhibits less clear
structure, with only some shells visible. In the outer regions, the
particles stripped from currently surviving objects are also
visible. These correspond to `streams' one would observe stemming from
dwarfs like Sagittarius. We note that these are less likely to
contribute to the fluctuations in the log-slope profile, as in
computing these profiles we take the angular average for a given bin
in distance. 

In the log-slope profile of the stars , several features are visible,
with log-slope values $<-6$. The dark matter, on the other hand, is
smoother than the stars on this diagram. Some shells can be seen, but
these are hard to distinguish from the background. In the log-slope
profile, one can clearly see the splashback radius just outside
$R_{200,\rm m}$ in the quiet example, and slightly inside $R_{200,\rm m}$ in
the active example (where the log-slope drops to -4); this is
consistent with the results of \cite{diemer_and_kravtsov}. Numerous
other minima in the log-slope of the dark matter can also be seen,
which seem to coincide roughly with the features in the stars,
interestingly including the splashback feature.

The similarities in the orbital properties of the dark matter and the
stars can be seen more clearly if we only select the dark matter
coming from stripped luminous haloes (which,  we note, is comparable in
amount to the smooth component, though the latter is dominant in
the outer regions). This is shown in the middle left panels of
Figs.~\ref{fig5} and \ref{fig6}, where it is clear that `shells' of
dark matter closely follow those of the stars and the features in the
log-slope of the density profile are significantly more pronounced.

As in \cite{edgeofthegalaxy}, we see a pronounced feature in the
log-slope of the stars at $\sim$0.6R$_{200,m}$, for both active and
quiet examples. In fact, several features are seen of comparable
log-slope. We note that since \cite{edgeofthegalaxy} analyzed the
lower-resolution version of the APOSTLE simulations, some of the drops
in the log-slope would likely combine into a single feature in their
analysis. Likewise, if we reduce the number of radial bins, nearby
log-slope features can merge into one, while varying the number of
angular bins can result in different steepness of the log-slope
caustics.

We now explore whether any of the log-slope features are common among
our samples of active and quiet Milky Way analogues. For this, we
stack the haloes in each subsample, weighting each particle by the
inverse of the total stellar/dark matter halo mass within $R_{200,\rm m}$
of the halo. We additionally normalize the radial velocity by the
value of the circular velocity at R$_{200,\rm m}$, to give equal weight to
each Milky Way analogue. We then compute the log slope. We find the
steepest feature in the log-slope of the stars between
0.5-0.6$R_{200,\rm m}$ for the quiet sample, and at 0.4$R_{200,\rm m}$ for the
active sample. Note that this reflects the radii at which the
contributions of disrupted luminous haloes are important in the two
samples.

\begin{figure}
    \centering
 
    \includegraphics[width=\columnwidth]{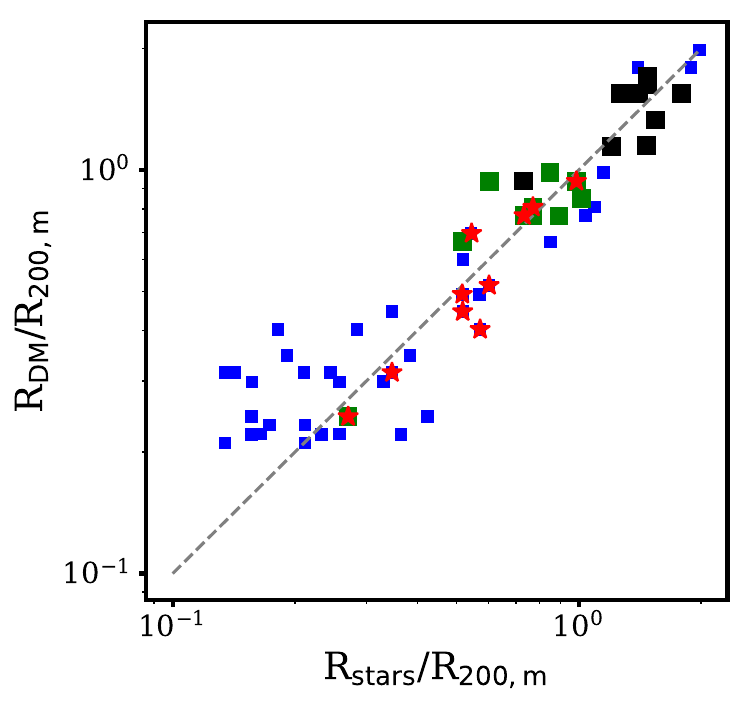} \\

    \caption{The minima in the log-slope of the density profile of the
      stars and the dark matter. For each minimum found in the
      log-slope of the stars, we find the nearest minimum in the
      log-slope of the dark matter density. The blue squares show all
      minima. The red stars show the steepest minima identified for
      each Milky Way / M31 analogue in the sample. The black squares
      show the closest stellar minima to the dark matter splashback
      radius. The green squares show the ``second caustic'' in the
      dark matter -- the minimum closest to the splashback radius.}
    \label{fig7}
\end{figure}

In Fig.~\ref{fig7} we demonstrate the similarity between the local
minima in the log-slope profile of the stars and the dark matter. For
each minimum in the stars, we find the nearest minimum in the dark
matter (blue squares). We see that the minima in the log-slope of the
stars often have a nearby minima in the log-slope of the dark
matter. Equivalently, the (over/under)densities in the stellar
distribution can be associated with (over/under)densities in the dark
matter. We do, however, see some differences at small radii. This is a
by-product of the noise in the distribution of the halo stars,
compared to a much smoother distribution of the dark matter in the
inner regions, given the binning we use to compute the profiles.

Red star symbols show the location of the steepest drop in the
log-slope of the stars. The black squares show the splashback radius
identified in the log-slope profile of the dark matter. The green
squares show the nearest minimum in the dark matter to the left of the
splashback radius (i.e. ``the second caustic''). One can see that this
is often not the steepest drop in the log-slope of the stars (i.e. the
"edge" of the galaxy, as defined by the steepest drop in the stellar
density log-slope, does not necessarily coincide with the second dark
matter caustic). Five out of 10 Milky Way~/~M31 analogues have their
steepest density log-slope between 0.5-0.6 $R_{200,\rm m}$, two are
located closer to $0.3R_{200,\rm m}$ and 3 above $0.7R_{200,\rm m}$. We have
previously shown that some of these discrepancies are due to the
accretion histories of these galaxies (i.e. active or quiet) . We will
examine this point further in the next section, when we will discuss
the origin of the shell features in the $v_r-D$ space.

\section{Formation of caustics}
\label{formationofcaustics}

\begin{figure*}
    \centering
    \begin{multicols}{3}
    \includegraphics[width =1.1 \columnwidth]{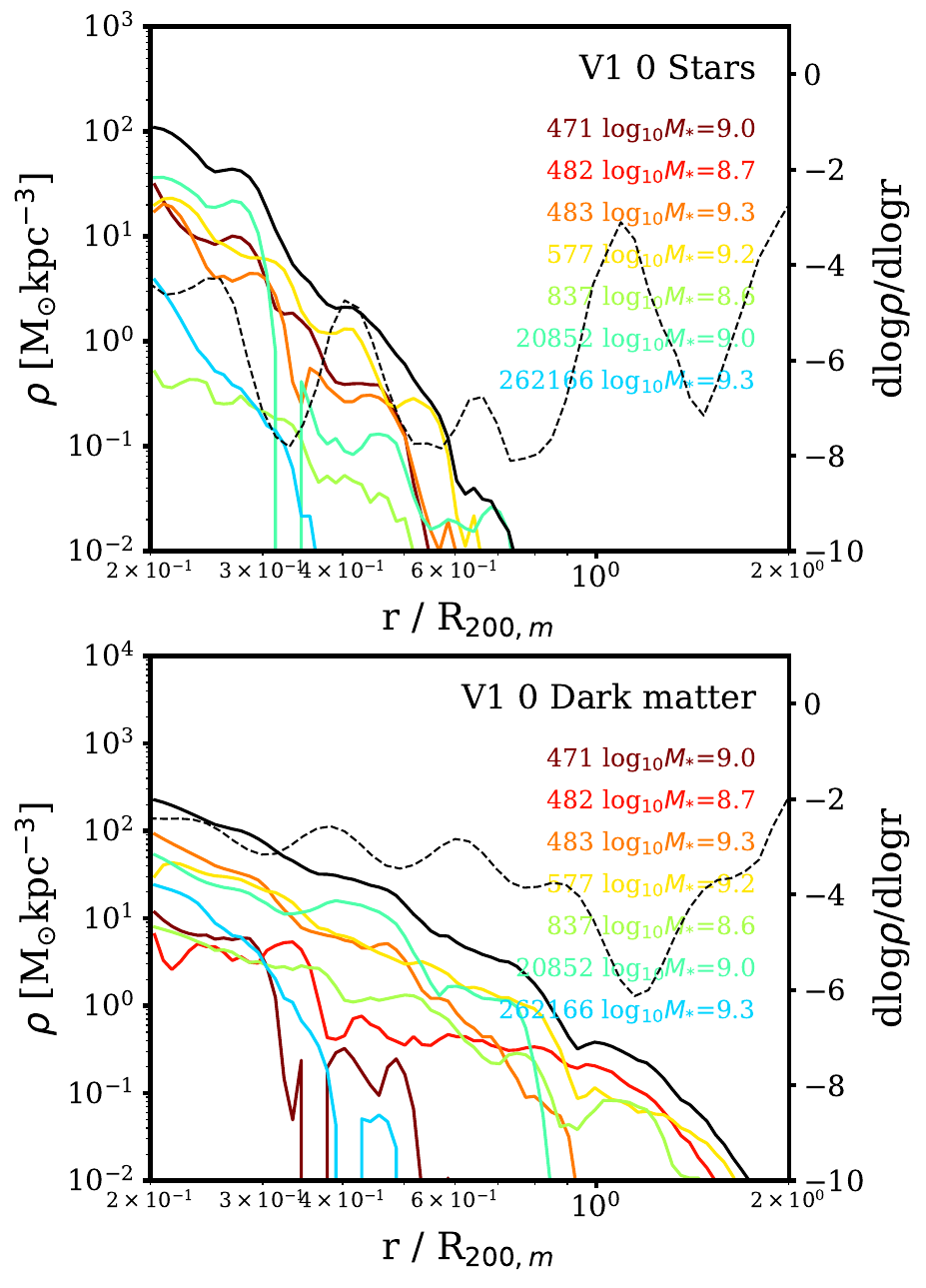} \\
    \includegraphics[width = 0.98\columnwidth]{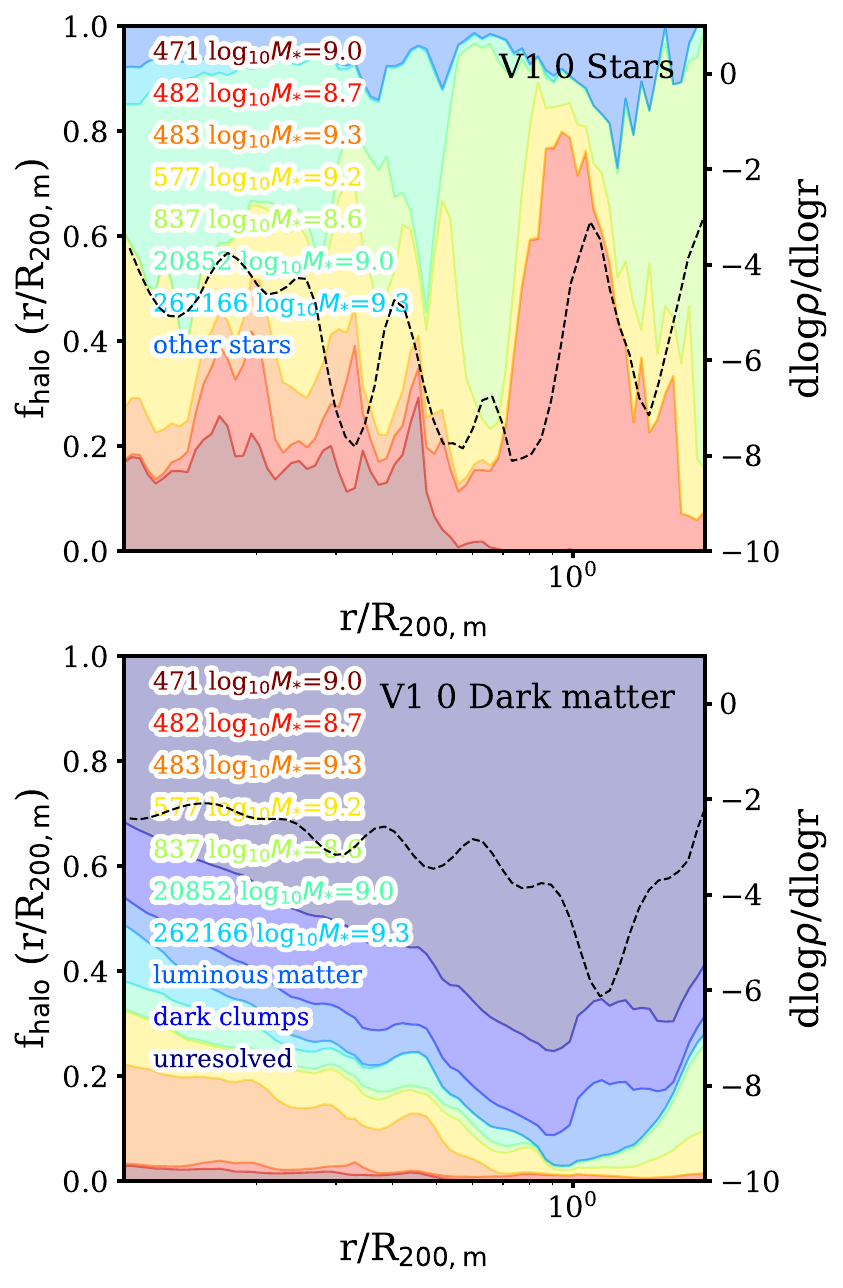} \\
     \includegraphics[width = 0.98\columnwidth]{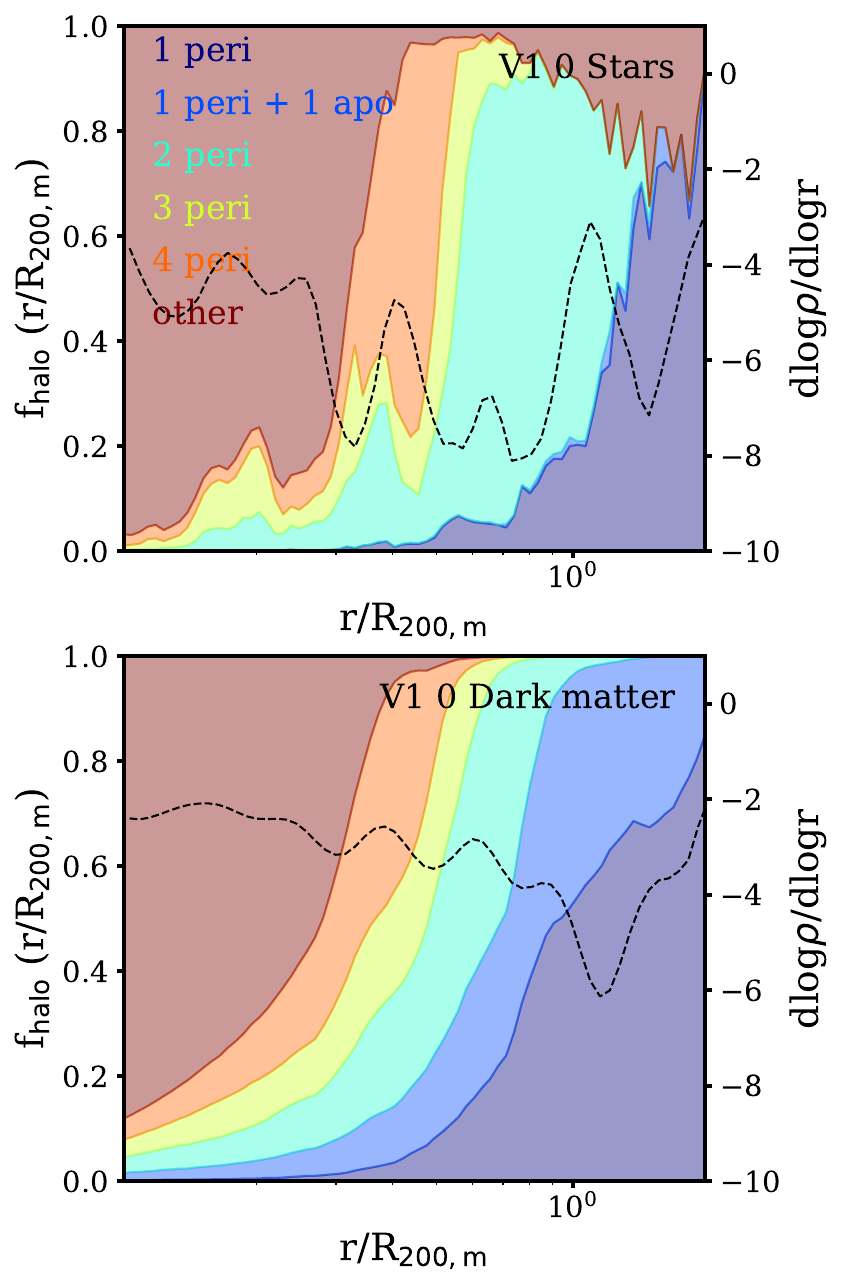}\\
    
    \end{multicols}
    \caption{\textit{Left:} density profiles of particles stripped from the biggest contributors to the stellar halo of Milky Way analogue \textbf{V1 0}. The peak stellar masses of each contributing dwarf are shown in the top right, together with their {\sc HBT+} identifying number. The black solid line is the combined density at each radius and the dotted line is the log-slope profile (with the scale shown on the right y-axis). \textit{Middle:} the radial composition of the stellar and dark matter haloes at each radius. The contributions are split into the most important contributors to the stellar halo, other dwarfs, dark haloes, and the `smooth' component. \textit{Right:} The radial composition of the halo at each radius, split into the number of pericentres that the particles have undergone since infall. The steepest drop in the log-slope of the stellar density occurs at the boundary of the 2 and 3-pericentre material. The splashback radius of the dark matter is located roughly where the material has completed half of an orbit (1 pericentre after infall and one consequent apocentre). }
    \label{fig8}
\end{figure*}

\subsection{Dark matter and stars stripped from infalling dwarfs}

We have so far established that the stars and dark matter stripped
from dwarf galaxies get deposited in the halo, forming `shells' in
$v_r-D$ space. In particular, stars and dark matter piling up close to
their apocentres cause radial overdensities, leading to features in
log-slope of the density profile. In this Section, we examine how the
stars and the dark matter stripped from dwarf galaxies infalling into
the Milky Way~/~M31 analogues are distributed in the halo. We will
focus on the particular example of a quiet galaxy, {\sc V1 0}, which
we have previously shown on the left panel of Fig.~\ref{fig6}. For
this analogue, we identify the biggest contributors to the stellar
halo, both disrupted and surviving, and plot the histogram of the
stripped particles' locations within the halo. This is shown on the
left of Fig.~\ref{fig8}. Different contributors are identified with
different colours, ordered by the stellar mass contributed to the
halo. Note that this is not necessarily reflected by the peak stellar
mass and that the greatest contributors of stars are not
necessarily the greatest contributors of dark matter.

We see a number of interesting features. First, we are able to
determine which dwarfs cause the `overdensities', corresponding to
particles piling up at the apocentres, and how these lead to the
fluctuations in the log-slope of the stellar density (black solid line
compared to black dotted line). Secondly, we see that the coincident
minima in the log-slopes of the stars and the dark matter are not
necessarily caused by the same dwarf galaxies. For example, the
minimum in the stars at $\sim0.8R_{200,\rm m}$ is caused by the stars
stripped from dwarf {\sc 837}, whereas the corresponding feature in
the dark matter seems to be due to some combination of {\sc 483, 577}
and {\sc 837}. We also see that each dwarf contributes substantial
amounts of dark matter at each `peak' (likely, subsequent apocentres
of stripped particles as the dwarf sinks), with some increase in
contribution towards the centre. At the same time, the stars are
stripped in small amounts at the outskirts of the halo and
substantially more towards the centre. Moreover, `peaks' of stripped
stellar and dark matter particles from the same dwarf galaxy do not
appear to always align. This may be one of the reasons for the offsets
observed between corresponding log-slope minima in the stars and the
dark matter: overlapping contributions from various dwarfs and the
differences in the stellar and dark matter stripping
\citep{libeskind}.

In the middle panel of Fig.~\ref{fig8}, we show the radial composition
of the stellar and dark matter haloes in cumulative form, focusing on
the contributions from the main stellar contributors, but showing in
addition the contribution from smaller luminous and dark haloes and
the smooth component. For the stellar halo in particular, it is clear
that our interpretation of the role of the top 4 mergers in creating
the caustics in the log-slope density profile is correct. For the dark
matter, the picture is somewhat more complex, as the four main stellar
contributors account for no more than 15~per~cent of the dark matter
halo out to $R_{200,\rm m}$; this figure primarily captures the radial
contribution of each component rather than the overall
density. However, one can see that of all components, it is the
contributions of dwarfs {\sc 577} and {\sc 483} that show oscillatory
behaviour that can be associated with variations in the log-slope. It
is remarkable that objects that contribute no more than 10-20~per~cent
of the dark matter halo out to $R_{200,\rm m}$ can cause such strong
variations in the density slope, which are also observable in the
stellar distribution.

\subsection{Completed pericentres }

It is tempting to associate the consecutive `peaks' in the locations
of particles of a given dwarf galaxy to the streams stripped off 
as the dwarf sinks to the centre of the host galaxy due to
dynamical friction. As particles are primarily stripped near 
pericentre, in this section we explore how the number of pericentres
completed by the particles in the halo relates to the particle spatial
distribution, and thus to variations in the log-slope of the
density profile \citep{diemer_pericentres}.

\subsubsection{Counting pericentres}

We find the number of pericentres that a particle has completed by counting
the number of times that its radial velocity, v$_r$, has changed sign. We also
require that an apocentre count can only follow a pericentre and
vice-versa. There are, however, some caveats to this method. Firstly,
we are largely limited by the frequency of output times in our
simulations, which decrease with decreasing redshift. If a particle
has completed an orbit and the orbital time is shorter than the time
interval between two simulation outputs, a pericentre cannot be
counted. This is a major problem of this method. However, as we show
in the following, the particles which make up the outer regions of the
halo have long orbital time periods and would have
completed typically no more than 4 full orbits during the dynamical
time of the Milky Way analogue. These particles are immune to the
limitation of infrequent simulation time outputs.

Another limitation is that subhaloes can interact with each other
within a host galaxy, sometimes flipping the sign of $v_r$ with
respect of the Milky Way.  We avoid this problem issue by requiring
that a pericentre or an apocentre of an orbit be counted only if the
change in the sign of $v_r$ lasts longer than one simulation
output. Again, this can undercount the number of pericentres or
apocentres for orbits with short time periods, but has no effect on
orbits with long orbital times which are of interest here.

\begin{figure*}
    \centering
    \includegraphics[width = 2\columnwidth]{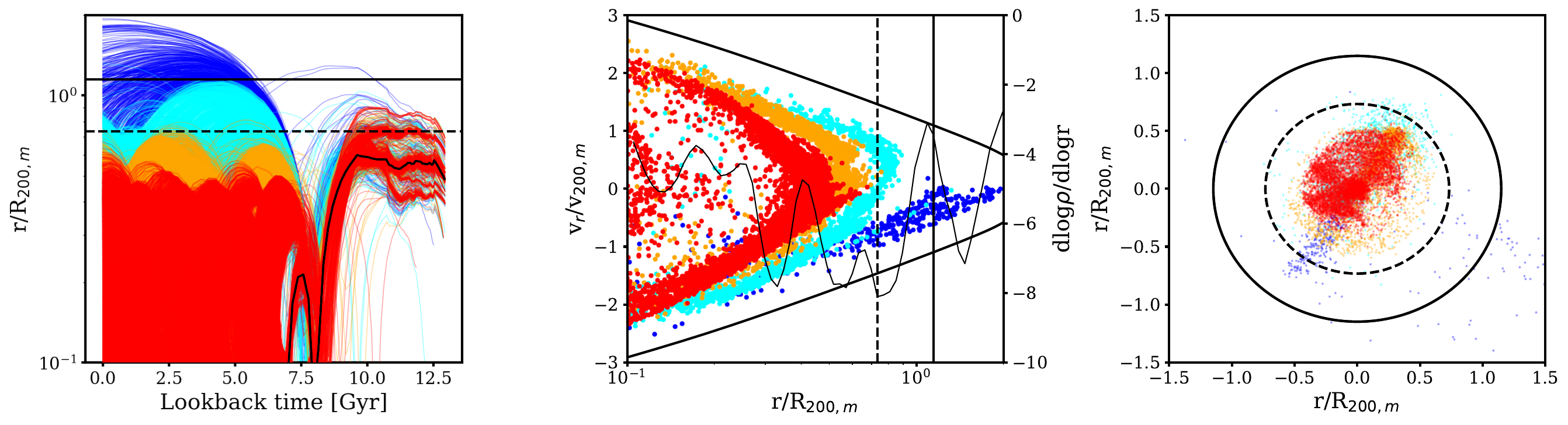}
    \includegraphics[width = 2\columnwidth]{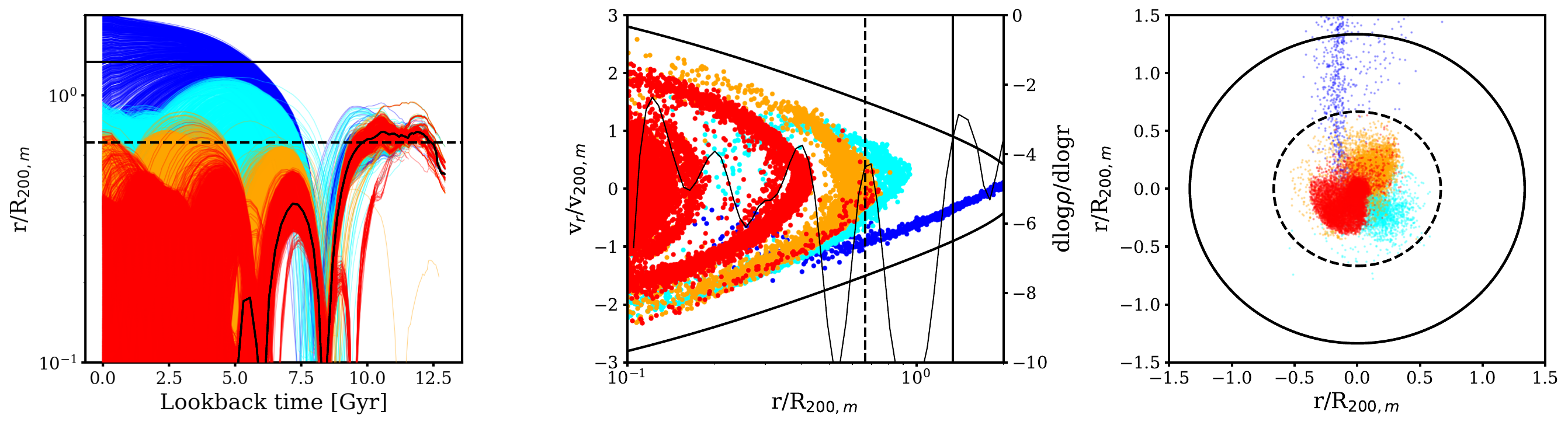}
    \caption{Examples of the stripping of massive dwarfs on more
      radial (top) and more tangential (bottom) initial orbits,
      leading to the formation of shell-like and umbrella-like
      features, respectively. \textit{Left:} positions as a function
      of time of stars that have undergone 1 (blue), 2 (cyan), 3
      (orange) and 4 (red) pericentres; the solid line tracks the
      orbit of the infalling dwarf galaxy. The dashed line is the
      location of the steepest drop in the log-slope of the stellar
      density profile. The black line is the splashback radius. The
      positions on the y-axis are in physical
      coordinates. \textit{Center:} phase-space diagram showing the
      stars that have completed each number of pericentres. The thin
      black curve is the log-slope of the density profile.
      \textit{Right:} x-y positions of the particles in the x-y
      coordinates of the simulation box.}
    \label{fig9}
\end{figure*}

\subsubsection{Stellar and dark matter halo split by number of pericentres}

On the right panels of Fig.~\ref{fig8}, we show the radial
contribution of particles with consecutive numbers of completed
pericentres for the stars (top) an dark matter (bottom). It is
remarkable to see the differences between the two. The stars have a
distinctly clumpy distribution, while the dark matter resembles almost
evenly spatially distributed `shells' of matter on consecutive
pericentres. In both cases, however, particles with more completed
pericentres dominate at smaller radii and particles with fewer
pericentres dominate at the outskirts.

A clear distinction between stars and dark matter in this case is the 
presence of dark matter particles that have had one pericentre and one
apocentre (lighter blue). These particles effectively define the
splashback radius of the halo and there are almost none visible 
in the stars. This likely reflects the differences in the
build-up of the two types of haloes. Dark matter includes the smooth
component as well as dark and luminous subhaloes which altogether
dominate the outer halo (see centre of Fig.~\ref{fig8}). The stellar
halo, on the other hand, is dominated by the debris from a few past
mergers in these regions, which have almost no stars with apocentres
reaching $\sim R_{200,m}$ (left of Fig.~\ref{fig8}).

The `edge' of the galaxy, as defined by the steepest drop in the
log-slope of the stellar density distribution, is coincident with the
region where the three-pericentre material dominates, while across the
two steepest log-slope drops, the 2-pericentre material is dominant
overall. The latter could be connected to dwarfs {\sc 577} and {\sc
  837} contributing to the stellar halo (see centre of
Fig.~\ref{fig8}).
 
\subsection{Examples of past major mergers}

\begin{figure}
    \centering
    \includegraphics[width = \columnwidth]{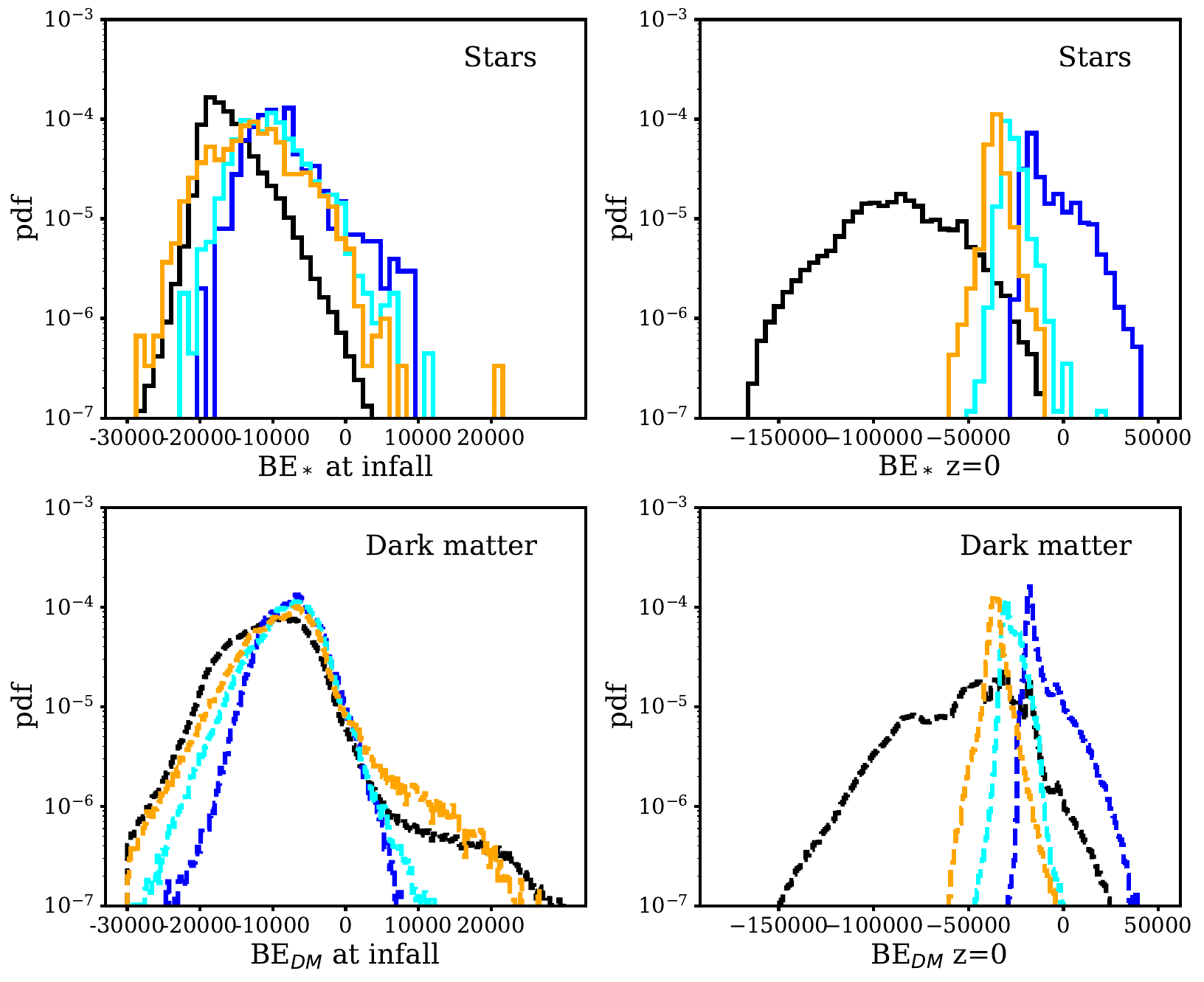}
   
    \caption{The binding energy of the star (top) and dark matter
      (bottom) particles of the merger shown in the top panel of
      Fig.~\ref{fig10}. The binding energy at the time of infall is
      shown on the left and the binding energy at $z=0$, with respect
      to the Milky Way analogue , is shown on the right. The blue,
      cyan and orange colours represent particles that have undergone
      1, 2 and 3 pericentres, respectively. The black lines show 
      all the particles that were bound to the dwarf at infall (i.e. when
      they crossed $R_{200,m}$).}
    \label{fig10}
\end{figure}

We now examine in more detail the processes that lead to the formation
of shells in the $v_r-D$ diagram. First, we identify dwarfs the which
contributed the most stars to the stellar halo of the Milky Way/M31 analogue. We
then follow the history of these objects -- their infall into the
Milky Way~/~M31 analogues and the stripping of their stars and dark
matter.

Figure~\ref{fig9} demonstrates the history of two dwarfs that merged
into two different Milky Way~/~M31 analogues. In the upper panel, the
infalling dwarf enters the halo on a very radial orbit, reaches its
first pericentre at $\sim8$~Gyr and has one further apocentre before
merging with the host halo. The orbital apocentre can be seen to be
rapidly damped by the effects of dynamical friction
\citep{amorisco_halo}. The colours in this plot denote the number of
pericentres that the particles have gone through by $z=0$, with blue
showing 1 pericentre and red 4 pericentres. We do not include
particles with more pericentres in this figure.

Firstly, one can see that the material located in the outer stellar
halo has been stripped primarily at the first pericentre of the
dwarf's orbit. From there, the stellar particles are dumped on a wide
range of orbits, showing a spread in orbital energies that leads to some
particles having  fewer pericentres than others. Those with longer
orbital times are the particles making up the outer regions of the
stellar halo. In the middle panel of Fig.~\ref{fig9}, it can be seen
that the `shells' in the $v_r-D$ diagram, corresponding to particles
with varying numbers of pericentres coincide with the minima in the
log-slope of the density. The particles that have only had one
pericentre (the first pericentre of the merging dwarf) are ejected in
some cases beyond the splashback radius of the halo (solid black
line) - some can become unbound. In the $v_r-D$ diagram,
these particles look as if they are being accreted onto the halo, with
generally negative radial velocities, while in the X-Y diagram it can
be seen that they form a kind of a `jet' from the centre of the
halo. These are akin to Gaia-Enceladus ``arches'' seen in the Toomre diagram \citep{koppelman, naidu_lmm}, which likely originate from stars on prograde orbits within the disk of the infalling dwarf or stars belonging to the dwarf's extended stellar halo.  These stellar particles, and their apocentres, could be the best
stellar tracer of the splashback radius of the halo (provided the halo
has not significantly grown since the merger). The particles which
have undergone at least two pericentres, and were originally dumped
with lower orbital energies, are the ones that define the steepest drop
in the log-slope of the stars. These are clearly on bound orbits
within the halo and define the stellar halo `edge', as proposed by
\citet{edgeofthegalaxy}.

We observe a similar behaviour in the merger shown in the lower panel
of Fig.~\ref{fig9}. In this case, however, the merger comes in on a
more tangential orbit. The dwarf is able to complete two full orbits
before effectively merging with the host halo. The increased
circularity of the orbit is also evident in the $v_r-D$ diagram, where
the shells of particles are clearly more circular, compared to the
`sharper' shells of a merger on a more radial orbit. The X-Y plot on
the bottom right of Fig.~\ref{fig9} demonstrates a distinct `umbrella'
shape, characteristic of mergers with higher angular momentum
\citep{martinez-delgado}. A `jet' of one-pericentre stars is visible
once again and extends out to beyond the splashback radius of the
halo. Note that the spread in orbital phases means that the `jet'
bends over, making a loop. Since the particles spend more time near
apocentre, the density is enhanced in the outer radii.

In Fig.~\ref{fig10} we show the binding energies of particles that
have undergone 1, 2 and 3 pericentres within the Milky Way analogue. In
the top panel, we show the binding energy with respect to the dwarf
galaxy at the time of infall (left) and with respect to the host halo
at $z=0$ (right). We show corresponding properties for the dark matter
at the bottom. It can be seen that the particles with successively
smaller number of completed pericentres were less bound in the dwarf
galaxy at infall. In the dark matter, one can see already a tail of
unbound particles at infall, while the stellar component is to a large
extent still bound. This confirms that the dark matter is stripped
earlier and more effectively than the stars. At the same time,
comparing the binding energy distributions within the Milky Way
analogue at $z=0$, one can see that the binding energies of the stars
and the dark matter with the same number of pericentres are remarkably
similar. As the binding energy is, in effect, the total energy of the
orbit, it best traces the apocentre of the orbit. This figure thus
reinforces the idea that the orbits of stellar particles stripped from
a dwarf follow the orbits of dark matter particles of the same energy
and have similar phase-space features. This also suggests that 
semi-analytical dark matter particle tagging techniques can give
faithful representations of the stellar distributions
\citep{bullock_and_johnston, gomez_aquarius, cooper_tagging}.

\subsection{The relation between past mergers and features in the log-slope of the halo density profile}

\begin{figure*}
    \centering
    \includegraphics[width =2\columnwidth]{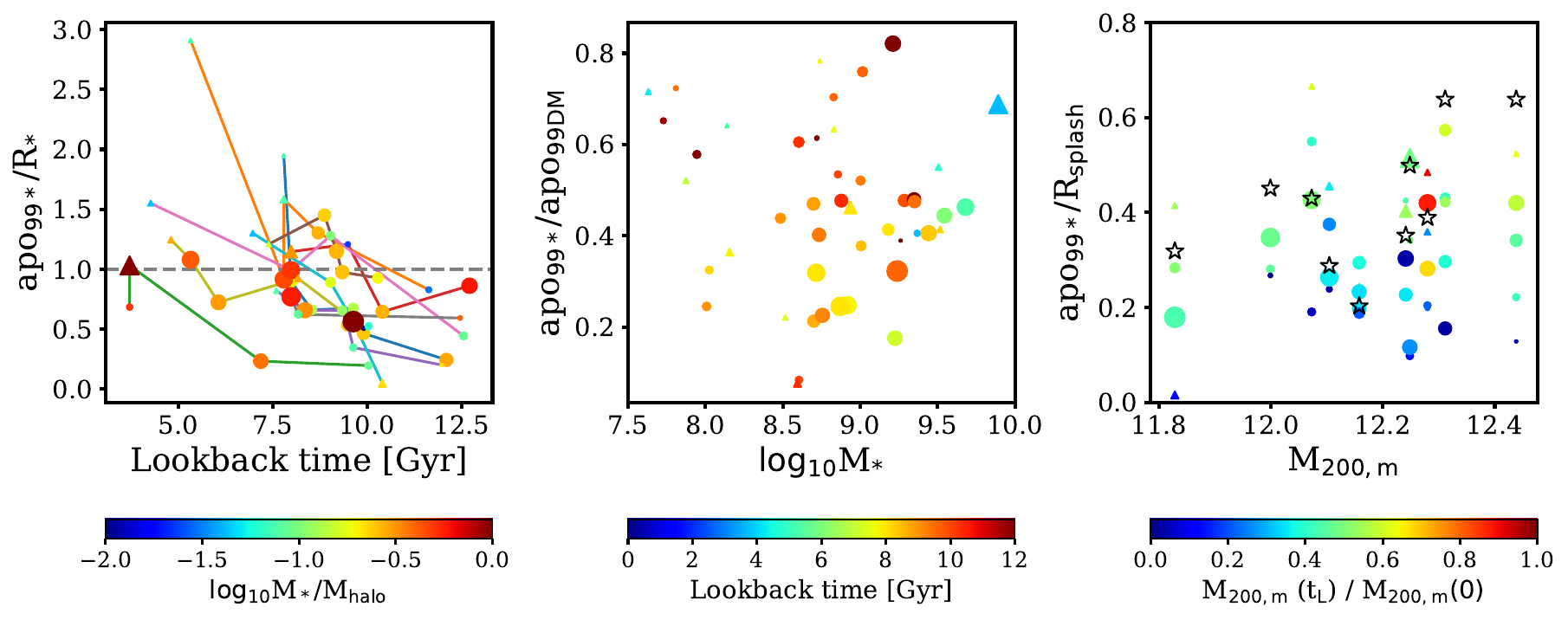}
    \caption{\textit{Left:} merger histories of the 10 Milky Way~/~M31 analogues. The horizontal axis is the lookback time of the merger (defined as the first pericentre), while the vertical axis gives the location of the 99$^{\rm th}$ percentile of the stellar particle apocentres of each merger event with respect to the location of the steepest drop in the log-slope of the stellar density profile, ${\rm apo_{99}/R_*}$. The points are coloured by the ratio of the dwarf stellar mass to the present-day Milky Way~/~M31 analogue stellar halo mass. Circles show fully disrupted dwarfs and triangles show dwarfs that have not been disrupted. The size of the points reflects the fraction of the stellar halo contributed by each dwarf. Lines of different colours represent individual Milky Way~/~M31 analogues. \textit{Middle:} ratio of the 99$^{\rm th}$ percentiles of stellar and dark matter particle apocentres for dwarfs of each peak stellar mass. The points are coloured by the lookback time of the merger. \textit {Right:} the location of particle apocentres with respect to the splashback radius, ${\rm apo_{99}/R_{splash}}$, as a function of the total mass within $R_{200,\rm m}$. The points are coloured by the fraction of the present-day $M_{200,\rm m}$ that was already in place at the time of the merger. The star symbols show the location of the steepest drop in the log-slope of the stellar density.}
    \label{fig11}
\end{figure*}

\begin{figure*}
    \centering
    \includegraphics[width = 2\columnwidth]{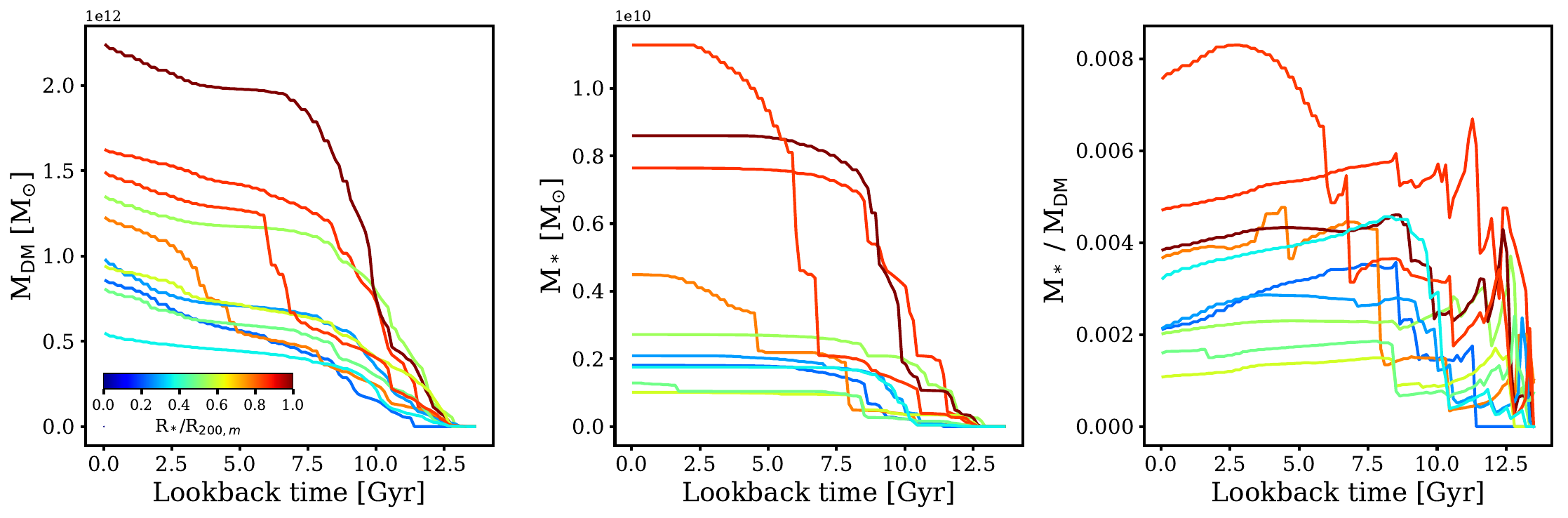}
   
    \caption{The assembly histories of the dark matter halo (left),
      the accreted stellar halo (centre) and the historical stellar
      mass-to-halo mass ratio of the Milky Way / M31 analogues
      (right). All are defined within $R_{200,\rm m}$. The lines are
      coloured by the location of the steepest drop in the log-slope
      profile of the stars, $R_*/R_{200,\rm m}$. Note that the low
      stellar-to-dark matter mass ratio does not imply a more embedded
      stellar halo.  }
    \label{fig12}
\end{figure*}

We now examine in more detail the relation between particle apocentres and the ``edge'' of the stellar halo, defined here as the location of the steepest drop in the log-slope of the stars. On the left panel of Fig.~\ref{fig11}, we display the merger histories of the 10 Milky Way~/~M31 analogues (identified with lines of different colours). The vertical location of each point represents the 99$^{\rm th}$ percentile of the apocentres of stars stripped from each dwarf. The dashed line marks the location of the steepest caustic. The sizes of the points reflect the fraction of the halo that each dwarf contributes, while  the colours reflect the stellar masses of each dwarf. In this figure we only include objects that contribute at least 1~percent of the accreted stellar halo mass. Several trends are visible. Firstly, the oldest mergers tend to have smaller particle apocentres, and so they do not typically define the ``edge'' of the stellar halo. We can also see that the oldest mergers are typically less massive and thus contribute a smaller fraction of the halo. More recent mergers, on the other hand, have stripped off particles that reach successively larger radii. In each case, we can identify the mergers which most likely define the galaxy ``edge''. These are typically dwarfs that fall in later and are, on average, more massive than those that came in earlier. As such, the stellar halo ``edge'' is caused by the apocentre pile-ups of the stars stripped in the last big merger. 

We now seek to establish the conditions that determine our ability to use stellar tracers to map dark matter. We have previously seen that log-slope features can be identified in both components at roughly similar locations;  however the stripping of the dark matter is more efficient than that of the stars and thus the ability of the stars to trace the dark matter in the halo will depend on the efficiency of the stripping and the similarity of the velocity distributions of stars and dark matter in the infalling dwarf. In the middle panel of Fig.~\ref{fig11}, we show the stellar mass of a dwarf, which is related to the stellar velocity dispersion, and the ratio of maximum apocentres of the stripped stellar and dark matter particles. It can be seen that dwarfs that have more stellar mass -- and thus higher velocity dispersion -- have stripped stars that follow the stripped dark matter orbits more closely. This relation shows scatter, which appears to be related to the time of the merger. For a given stellar mass, the stripped stars trace the dark matter better for early mergers. This suggests that tidal stripping may be more efficient early on in the history of the Milky Way and could, in part, be due to a lower concentration of the dwarfs at higher redshift.

Given the differences between the stripping of stellar and dark matter particles, it interesting to ask if  there is a relation between the edge of the stellar halo and the location of the splashback radius.  On the right of Fig.~\ref{fig11}, we show the ratio of ``edges'' of merger debris to the splashback radius as a function of M$_{200,\rm m}$. The points are coloured by the fraction of total present-day mass within $R_{200,\rm m}$ that was in place at the time of the merger (this includes the merging dwarf). We can see that the more massive Milky Way~/~M31 analogues typically have galaxy ``edges'' that extend closer to the splashback radius, although there is some scatter. We note that the ``edges'' do not necessarily correspond to the furthest apocentres of the  merger debris in some cases (see Fig.~\ref{fig7}), though the stellar component beyond the edge is typically rather diffuse.  

We find that the ``edges'' of our Milky Way analogues range between 0.2-0.65 R$_{\rm splash}$. We can see that the cases where the ``edge'' of the halo is more embedded also corresponds to the cases where the halo has not assembled more than 50~percent of its final mass at the time when the largest contributors to the stellar halo merged with the main galaxy. These objects correspond to our `active' sample of Milky Way~/~M31 analogues. Overall, Fig.~\ref{fig11} suggests that there is no direct conversion between the galaxy ``edge'' and the splashback radius; relating these two properties may require knowledge of the recent growth history of the dark matter halo. We expand on this feature of some of our Milky Way/M31 analogues in the next section.

\subsection{What determines the ``edge'' of the galaxy?}

We now explore the main factors behind the location of the steepest drop in the log-slope of the density profile of the stars relative to $R_{200,\rm m}$. From past work on the splashback radius of the halo, some of the main factors were identified as the halo mass and the mass accretion rate of the halo, typically defined from a halo dynamical time of $\sim 8.5$~Gyr ago to the present day. While our analogue galaxies were selected to match the constraints for the Milky Way and M31, there is still a noticeable variety in stellar and dark matter halo masses. In Fig.~\ref{fig12}, we show the mass assembly histories of the dark matter halo (left) and the stellar halo (centre). The right panel shows the evolution of the stellar-to-dark matter halo mass ratio. The lines are coloured by the location of the galaxy `edge', defined by the location of the steepest caustic.

From the left panel of Fig.~\ref{fig12}, in agreement with the right panel of Fig.~\ref{fig11}, it is clear that the more massive dark matter haloes typically have their stellar halo edge further out relative to $R_{200,m}$, whereas the less massive haloes have a stellar halo edge that is more embedded. This suggests that dark matter halo mass is an important driver of the location of the stellar halo edge. Nevertheless, we also see a number of outliers -- for example, the fourth most massive halo has an edge at $\sim0.5 R_{200,\rm m}$, while the fifth most massive has it at $0.7 R_{200,\rm m}$. Galaxies with bigger stellar haloes also tend to have a higher stellar mass, though it does not appear to be the case that the stellar mass alone can explain the location of the halo edge (middle panel of Fig.~\ref{fig12}). 

To further investigate the source of this diversity, we look at the historical stellar-to-dark matter halo mass relation (right panel of Fig.~\ref{fig12}). We see that the more massive dark matter haloes also tend to have higher $M_*/M_{\rm DM}$, as expected \citep{behroozi, moster}. If the total stellar mass to dark mass in the infalling dwarf is relatively high, this could result in more stars being stripped earlier on after infall, allowing stars to trace the stripped dark matter further out in the halo and simultaneously `pushing' the stellar halo edge further out. 

However, it does not appear to be the case that the smaller values of $M_*/M_{\rm DM}$ lead to more embedded stellar haloes. For example, the Milky Way~/~M31 analogue with the lowest $M_*/M_{\rm DM}$ has its stellar halo edge at $\sim0.6R_{200,\rm m}$, while a halo with  $M_*/M_{\rm DM} \approx 0.003$ has its edge at $\sim0.3R_{200,\rm m}$. However, there is one feature that distinguishes the 'blue' curves from the `green': the rate of change of $M_*/M_{\rm DM}$, whereby haloes with more embedded haloes have $M_*/M_{\rm DM}$ declining more steeply with time. From the middle panel of Fig.~\ref{fig12}, it is clear that the stellar haloes are almost completely assembled $\sim8$~Gyr ago. This suggests that the main driver of the change in $M_*/M_{\rm DM}$ is a faster dark matter assembly. The growth of the dark matter halo in the last few gigayears leads to the increased concentration of the stellar halo within the dark halo, although the total halo mass is also an important factor.

The Milky Way~/~M31 analogues where the ``edge'' is more concentrated relative to $R_{200,\rm m}$ or the splashback radius belong to our `active' sample. Looking at the stack of this sample on the right of Fig.~\ref{fig5}, it is clear what causes this accelerated growth in the dark matter mass: these objects tend to have a massive satellite that deposits a large amount of the dark matter into the halo, but not so much in the stars. This is reminiscent of the Large Magellanic Cloud, which could have contributed a significant fraction of the total mass of the Milky Way, but has not experienced significant stripping of stars as it is likely on its first infall \citep{besla_lmc, conroy_lmc,petersen_lmc, lmc_mw_mass}. 

\section{Observational prospects}
\label{observations}

\begin{figure*}
    \centering
    \includegraphics[width =2\columnwidth]{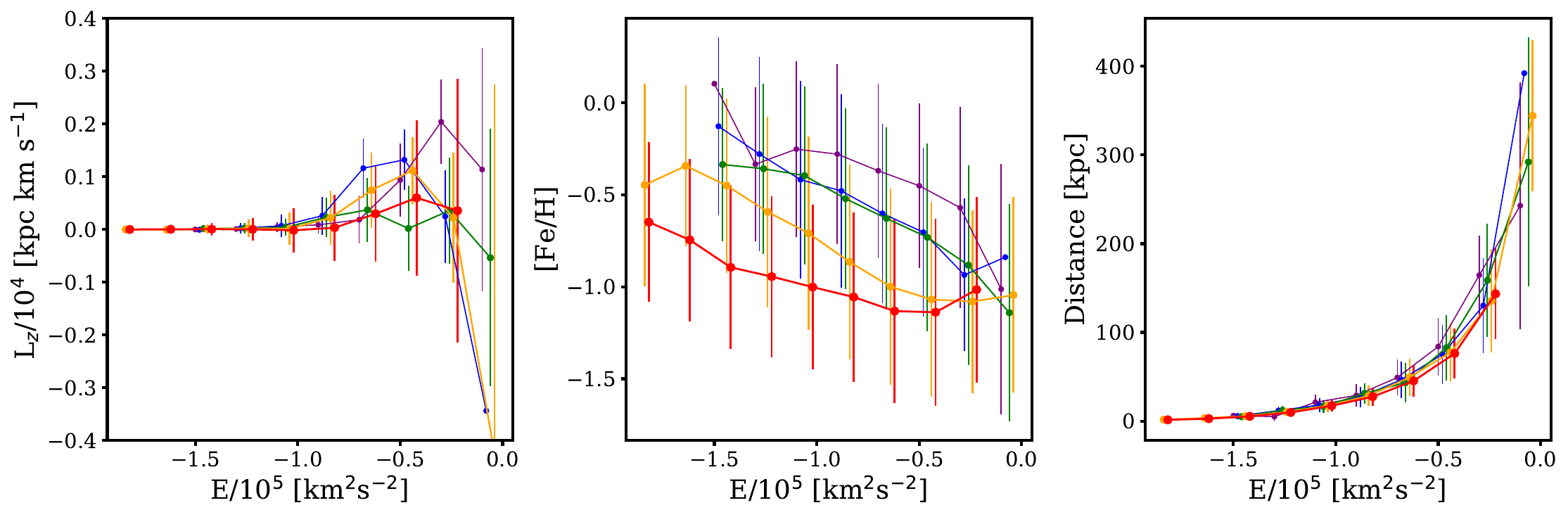}
    \caption{The properties of particles for the 5 most important
      contributors to the stellar halo of \textbf{V1 0} shown with
      different colours. The size of the points shows the
      fraction of the stellar halo made up by each
      dwarf. \textit{Left:} angular momentum of the particles as a
      function of orbital energy. \textit{Centre:} metallicity as a
      function of orbital energy, in solar units. \textit{Right:}
      distance from the Milky Way / M31 analogue centre as a function
      of energy.}
    \label{fig13}
\end{figure*}

\begin{figure*}
    \centering
    \includegraphics[width =1.5\columnwidth]{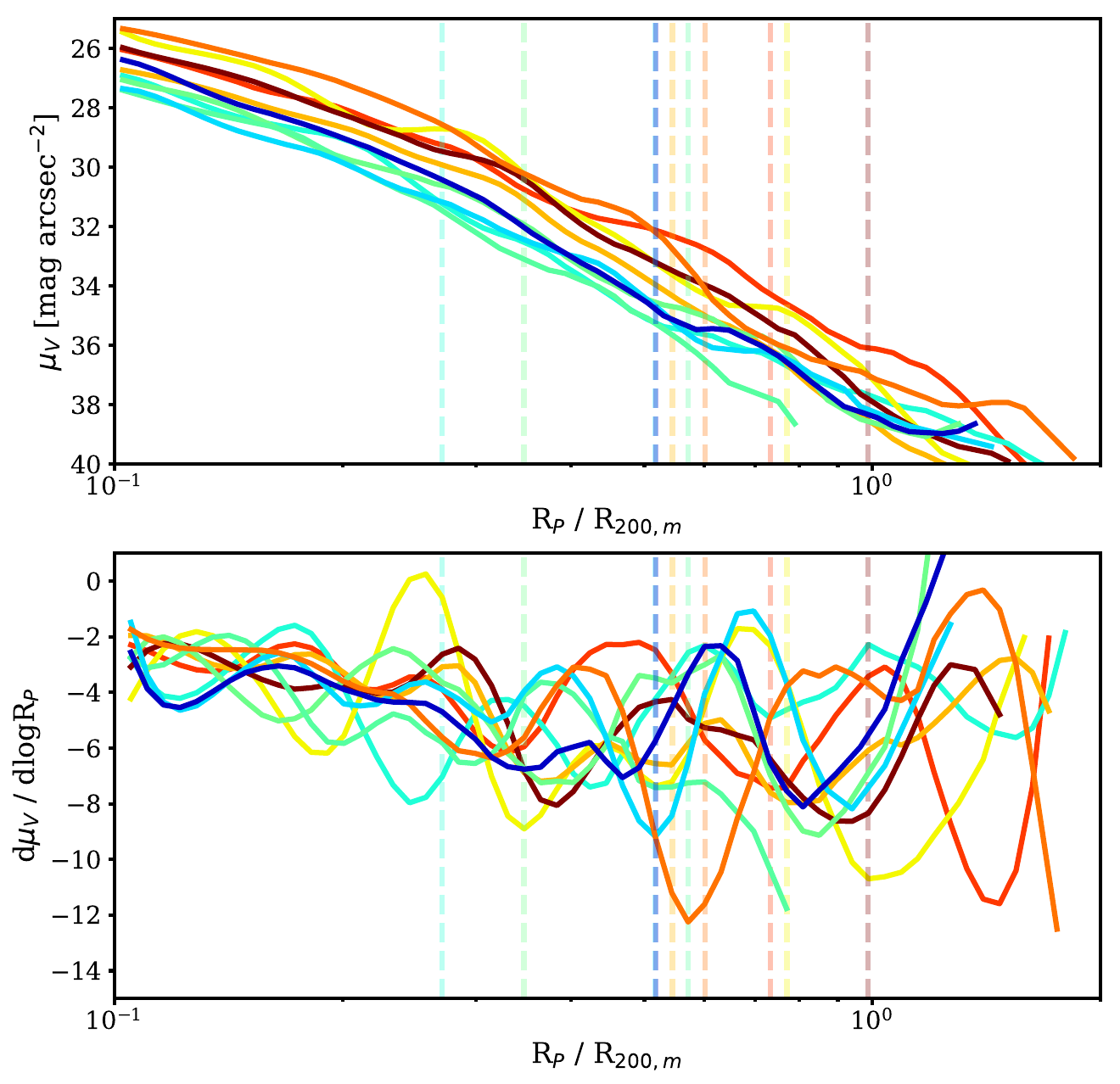}
    \caption{The stellar surface brightness profiles in the V-band
      (top) and the slope of the surface brightness (bottom) as a
      function of radius, normalized by $R_{200,\rm m}$. Each Milky
      Way~/~M31 analogue is labeled with a unique colour, reflecting
      the mass of the dark halo (blue is lower mass, and red is higher
      mass). The vertical dashed lines mark the location of the
      steepest caustic in 3D.  }
    \label{fig14}
\end{figure*}

In this section, we discuss the prospects for identifying the remnants
of the past mergers that contribute to the stellar halo through their
kinematic and chemical properties. We further explore whether halo
`edges' of external galaxies can be identified with deep photometry
and connected to the underlying dark matter halo and its
assembly history.

\subsection{Chemo-kinematic properties of past mergers}

In Fig.~\ref{fig13}, we select 5 important mergers in the history of a
quiet Milky Way analogue. These are identified with different colours
and the symbol size corresponds to the fraction of
stars they contribute to the stellar halo. It can be seen that these
different dwarf galaxies are indistinguishable in the energy-angular
momentum space at small radii, where most of the data are available
(the right panel of Fig.~\ref{fig13} shows the typical galactocentric
distance at each energy). Any significant deviations from the mean
energy and angular momentum can only be seen in the outer regions of
the halo, beyond $\sim 100$~kpc or so, where the stars have not yet
phase-mixed. However, the contributing dwarfs have somewhat different
metallicities. This suggests a way to distinguish past
mergers. Moreover, a metallicity gradient can be seen, whereby the
stars in the outer halo are more metal-poor than the stars in the
inner halo. The stars which are more bound also tend to have lower
metallicities, suggesting that the most bound stars come from the most
ancient mergers. While the stars are significantly phase-mixed in the
inner regions, making it difficult to distinguish different
progenitors through their kinematics, the metallicities can differ
sufficiently to tell the separate components apart. If the stellar
ages are taken into account, one could use the redshift-dependent
mass-metallicity relation to disentangle the different progenitors (see e.g. \citealt{monachesi_2019}).

Overall, our results suggest that detailed chemistry and, ideally,
stellar ages are required to disentangle the origins of individual
stars in the inner regions of the halo (see e.g. \citealt{naidu}), while
kinematics are sufficient to identify individual structures in the
outer halo. Note, however, that our results also suggest that the
stars stripped from the same dwarf can have prograde and retrograge
motions and in that case one must employ additional information to
avoid classifying these as separate structures
\citep{virgo_herc_common, kim_multiple, amarante}. These results are in agreement with the findings from the HESTIA simulations \citep{hestia_halo}, who show that the debris from a single merger occupies a wide area of $L_z - E$ space  that overlaps with other mergers, but also has several overdensities in this space.  The asymmetric and time-variable halo potential leads the location of the debris in the action-angle space to change over time. This work also found that enhanced star formation in the dwarf that is about to merge leads the ages of the most recently stripped stars to correspond to merger time. This feature may be used to recover the Milky Way's assembly history.

\subsection{The surface brightness of the stellar halo edge}

We have so far shown that some of the most important mergers that
define the `edge' of the stellar halo of Milky Way-like galaxies leave
the most kinematically distinct traces in the outskirts of haloes,
beyond $\sim100$~kpc. Aside from kinematics, we have shown that the
stellar density profile also shows variations due to particle
`pile-ups' at apocentre, which are associated with the dark matter
halo mass and assembly history.

What do the variations in the log-slope of the stellar density profile
mean in terms of surface brightness? To compute the AB V-band
magnitudes of the stellar particles in our simulations we use the {\sc
  fsps} software \citep{fsps1,fsps2}, where we adopt the Padova
stellar isochrone library \citep{padova1,padova2} and a
\cite{chabrier} initial mass function. For each stellar particle, we
provide its age and smoothed metallicity to the code. In the top panel
of Fig.~\ref{fig14}, we show the surface brightness profiles of our 10
Milky Way~/~M31 analogues, computed using an arbitrary projection. In
order to avoid contributions from the closest galaxy (as our systems
form a Local Group), we exclude all particles within $R_{200,\rm m}$ of
the companion halo. As for the 3D profiles, we compute the profiles in
75 radial bins and 11 angular bins, taking the median of the angular
bins at each radius. We smooth the profiles using the Savitzky-Golay
filter \citep{savitzky}. Each analogue is identified with a unique
colour. One can see clearly that the surface brightness profiles of
the haloes are not smooth, but exhibit variations similar to those
found by \cite{edgeofthegalaxy}. These variations are substantially
more pronounced in the computed surface brightness slope (bottom panel
of Fig.~\ref{fig14}).

The vertical dashed lines in Fig.~\ref{fig14} show the location of the
steepest drop in the log-slope of the 3D stellar density profile. This
is typically further than the nearest drop in the log-slope of the
surface brightness, as expected from projection effects. While the
location of the halo edge varies, as we have discussed previously, the
typical location, 0.5-0.6~$R_{200,\rm m}$, occurs at a V-band surface
brightness of between 31-36 mag~arcsec$^{-2}$, marginally achievable
with the Euclid Deep Survey \citep{euclid}, Hubble Ultra Deep
Field \citep{hubble_udf} and 10-year Legacy Survey of Space and Time \citep{lsst_deep}, albeit with typically smaller survey
areas. Targeted low-surface brightness surveys, like Dragonfly Nearby Galaxies Survey (DNGS) are already achieving comparable levels of surface brightness \citep{dragonfly}. An alternative is to stack images of Milky Way-mass galaxies \citep{stack_dsouza, stack_wang},
as has been done in the past in the search for splashback radius in
galaxy clusters \citep{splash_stack}.

\section{Summary and Conclusions}
\label{conclusions}

The Cold Dark Matter model implies hierarchical structure formation in
our Universe, where smaller structures form first and accumulate to
former larger ones. The model predicts the typical assembly
histories, with scatter, for galaxies such as our own Milky Way. Since
the presence of dark matter has been observed only indirectly, one has
to rely on the visible baryonic component to infer the total matter
content and the assembly history of galaxies like the Milky Way. The
emergence of cosmological hydrodynamic simulations has allowed us to
model the formation and evolution of galaxies within the $\Lambda$CDM
paradigm. These simulations make predictions for how the stellar halo
of the Milky Way has assembled through past accretion events and how
this relates to the assembly of the dark matter halo. In particular,
one can look for signatures in the stellar component that would reveal
the properties of the dark matter. In this work we have examined the
build-up of Milky Way-mass haloes in Local Group-like environments
from the APOSTLE suite of simulations. We have examined both stellar
and dark matter halo build-up through accretion. We find the following: \\

\noindent{\it i)} In the CDM paradigm
of hierarchical structure formation, large dark matter haloes are
built up through accumulation of smaller clumps \citep{frenk88}. By mass, subhaloes
that have hosted stars make up 30-40~per~cent of a galactic dark
matter halo, with the `smooth' halo component making up the majority
of the mass (35-40~per~cent). The smooth component is itself split
into particles that are not in bound structures as well as those in
haloes of mass below the resolution limit of our simulations (subhaloes < 10$^6$
M$\odot$, where the power-law form of the CDM mass function breaks
down.) The contributions of the dark and the luminous components are
quite similar and it is their relative sizes that determine the
degree to which the stars in the halo are able spatially to trace
the dark matter (the stellar mass -- halo mass relation). \\

\noindent{\it ii)} The accreted stellar halo of Milky Way-like galaxies is
primarily built up from disrupted dwarfs ($\sim85$~per~cent). Stars
stripped from surviving dwarfs typically make up 10-15~per~cent of the
stellar halo. It is typically 5-6 dwarfs with peak stellar mass of
$>10^9$M$_{\odot}$ that make up $\sim80$~per~cent of the stellar halo
in Milky Way-mass galaxies; of those the majority
are disrupted. \\

\noindent{\it iii)} We identify `active' and `quiet' Milky Way~/~M31 analogues
in our sample of 10 galaxies, in equal numbers. The main distinction
between the two is the relative contribution of surviving subhaloes and the distribution of their debris in the halo. This has to do with the order in which particles
are stripped from dwarf galaxies -- dark matter stripping occurs
before stripping of stars due to the more extended spatial
distribution of dark matter particles and their lower binding energies
\citep{libeskind}. We also find that the halo stars in `active'
galaxies are more centrally concentrated than in the quiet sample,
which overall results in more embedded stellar haloes. Active galaxies
also have a more significant dark matter contribution from disrupted
dark haloes in the outer regions. Radial accretion and stripping time
gradients suggest this is due to subhaloes that began to be stripped
prior to crossing $R_{200,\rm m}$ while in a group with larger haloes. \\

\noindent{\it iv)} The disruption of dwarf galaxies as they fall into the Milky
Way leaves structural imprints on the phase-space distribution. On a
$v_r - D$ diagram, this takes the form of shells of particles
following similar orbits in both stars and dark matter. Structures
seen in the stellar halo have corresponding structures in the dark
matter, although in the latter case they are ``smoothed out" due to
the dominance of the smooth dark matter component, particularly beyond
$0.1R_{200,\rm m}$ (smaller radii are not better places to look for such
structures due to increasingly phase-mixed material in those
regions). If the smooth component is removed from the analysis, the
$v_r-D$ structure of the dark matter halo in the outer regions follows
closely that of
the stars. \\

\noindent{\it v)} In agreement with \cite{edgeofthegalaxy}, we find that the
log-slope of the stellar halo density has a prominent trough at
$\sim0.5 - 0.6$R$_{200,\rm m}$, although the exact location varies,
corresponding to apocentre pile-ups of particles that have completed
two-three orbital pericentres since infall into the Milky Way~/~M31
analogues. However, in this work we have examined simulations with
higher mass resolution than \cite{edgeofthegalaxy}. As such, we were
able to detect additional features in the stellar and dark matter halo
density profiles, located closer in. Overall, we have found that even
out to the splashback radius, variations in the dark matter density
profile have corresponding variations in the density profile of the
stars. \\

\noindent{\it vi)} We examined the formation of the shells that lead to the
formation of the `edge' of the galaxy (i.e. the steepest drop in the
log-slope profile of the stars). We found that typically one or two
mergers deposit particles in the outer regions that lead to 
deviations of the halo density profile from smoothness. Contributions
from several important mergers can also add up to enhance variations
in the density profile. Since the stripping of the dark matter is more
efficient and more continuous along an orbit than for the stars, the
`peaks' in the radius of stripped dark matter particles can be offset
from those of the stars stripped from the same objects, leading to
slight differences in the locations of the density log-slope
minima. Additionally, the dark matter content at large radii, where we
expect to see the stellar halo edge, is dominated by the smooth
component which suppresses variations in the density profile of the
dark matter compared to corresponding features in the stars.\\

\noindent{\it vii)} We found a common behaviour in mergers that have
contributed the most to the stellar halo of Milky Way-like
galaxies. In agreement with previous work, we found that these objects
enter the halo on very radial orbits due to their large mass and
the undergo dynamical friction. The log-slope features that are
detected in the stellar halo outskirts correspond to particles that
were typically stripped at the first pericentre of the satellite's
orbit. The wide range in particle binding energies, particularly in
the most massive mergers, leads to a spread of particle apocentres and
orbital phases. Particles which were less bound within the infalling
dwarf galaxy end up on more energetic orbits within the Milky Way
after they are stripped and, consequently, complete fewer orbits by
the present day. Depending on the time of the merger, a `jet' of
1-pericentre stellar particles can be seen extending from the halo
centre and looping around. If the halo has not grown significantly
after the merger, these particles roughly trace the splashback radius
of the dark matter halo. The steepest trough in the log-slope of the
stellar density is the result of apocentre pile-ups of particles that
have completed 2-3 orbits, depending on the halo growth since the
merger. However, if the halo growth is particularly fast after the
merger, these relations can break down as the stellar halo is
dominated by particles that have not completed a full orbit. Although
we have not explicitly demonstrated the same features in the dark
matter, we have shown that stellar and dark matter particles with the
same number of completed pericentres have very similar distributions
of binding energies within the Milky Way analogues.\\

\noindent{\it viii)} We have looked into the specific histories of the dark
matter and stellar halo growth, and their relation, to establish the
main drivers of the location of the halo `edge'. We have found that
a large halo mass, which often corresponds to a large mass in the
stellar halo, leads to more extended haloes. This reflects the fact that 
these haloes experienced more massive major mergers that had higher
stellar mass contributions, allowing the stars to trace the dark
matter out to larger radii. At the same time, halo mass does not seem
to be the only important factor. Since the stellar haloes often
assembled earlier than the dark matter, the subsequent dark halo
growth leads to increased concentration of the stellar halo within the dark halo. This is
consistent with previous work exploring the halo splashback radius. In particular, we find stellar halo ``edges'' to be more embedded within the dark matter halo if the halo has recently accreted a massive satellite that contributes roughly 25~per~cent of the dark halo mass, while its stars have generally not been affected by tidal forces. \\

\noindent{\it ix)} We have examined the possibility of uncovering the important
mergers contributing to the build-up of the stellar halo. We have
found that the debris from these past mergers has a metallicity
gradient across the stellar halo, likely stemming from the metallicity
gradient within the objects themselves. More massive mergers typically
have particles with higher binding energies (lower orbital energies)
within the stellar halo and thus contribute more particles at small
radii (central $\sim 20$~kpc), making them more likely to be
detected. At the same time, across all radii, there is a large spread
in the metallicity distribution that may overlap with other
contributors. We have also examined the distributions of the orbital
energy and angular momentum that may help distinguish different
mergers kinematically. We found that at small radii (below
$\sim 50$~kpc) the orbits are rather similar, likely due to mixing in
a turbulent time-varying gravitational potential; however, at large
radii the particle orbits of stars from different progenitor dwarfs
become increasingly distinct from each other. This is likely where the
particles are still on the first or second pericentre of the orbit,
moving still somewhat coherently since the time they were stripped
(see, for example, the `jet' features in Fig.~\ref{fig9}). \\

\noindent{\it x)} We find that the `edge' of the stellar halo in Milky Way-like
galaxies typically corresponds to a surface brightness of 31-36
mag~arcsec$^{-2}$. Reaching this surface brightness limit is
marginally possible with existing, though coverage-limited, ultra deep
photometric surveys, like DNGS \citep{dragonfly}. Alternatively, one may stack multiple images of Milky Way-mass galaxies in search of the `edge' feature \citep{stack_dsouza, stack_wang}. \\

In this work, we have examined the predictions of the $\Lambda$CDM
model, together with the {\sc EAGLE} model of galaxy formation for the
assembly of the accreted stellar and dark matter halos of Milky Way~/~M31
analogues in Local Group-like environments. Some aspects of this work
may be sensitive to the assumed galaxy formation physics. For example,
the details of the stellar-halo mass relation may, to some extent,
alter the ability of stripped stars to track the dark matter. This
could also be affected by the sizes of the galaxies. We note, however,
that the galaxy formation model we have employed has been shown to
reproduce these galaxy scaling relations \citep{eagle1, eagle2, campbell}, with the largest uncertainties expected both above the Milky Way mass and in the regime
of the classical dwarf galaxies, in which the observed relations are
not well constrained.  We also do not expect small changes in the halo
mass threshold above which galaxy formation occurs to affect our results
significantly, since it is the most massive galaxies that contribute
the majority of stars in the outer halo.

One limitation of this work in the context of the direct comparison of simulated haloes to the Milky Way and M31 lies in the morphology, assembly history and mass of the stellar discs. APOSTLE halo pairs were selected to match the observational constraints on the separation of Milky Way and M31 and their total mass. This does not however guarantee comparable formation histories, only a similar present-day local environment. The main galaxies in APOSTLE are known to be less massive than the Milky Way and M31 \citep{fattahi}, which leads to reduced disruption of substructure by the potential of the disk \citep{richings}. We expect however, that this does not change our conclusions on the contribution of the most massive dwarfs to the stellar halo, as these objects quickly sink and merge with the host galaxy due to dynamical friction. In addition, the work of \citet{kelly_apostle_auriga} investigated the properties of disks in APOSTLE high-resolution volumes (which we use here) in comparison to the same galaxies simulated with the AURIGA model \citep{auriga} at the same mass resolution. These authors found that the discs of the main galaxies in APOSTLE were too thick, had a factor of $\sim2$ larger disc scale length and lacked the spiral arm and bar structures seen in AURIGA simulations, while the latter also formed a factor of $\sim2.5$ more stellar mass. The authors suggested that this stems from the difference in effective spatial resolution, which is higher in mesh-based codes compared to SPH codes. Previous works have also shown that the disc can bias the debris of satellites accreted on low-inclination orbits into the disc plane \citep{quinn_1993,read_2008}. This may imply that the orientation of the satellite debris is somewhat more isotropic in our simulations than in typical disc galaxies. Nevertheless, this is unlikely to affect our results regarding the origin of the `stellar splashback'. 

What is perhaps more concerning is the larger thickness of the discs of star-forming dwarf galaxies in APOSTLE compared to observations \citep{kylecurves}, because the higher velocity dispersion in the stellar disc may cause premature stripping of the stars and thus an enhancement of the mass of the stellar debris deposited in the outskirts of the Milky Way~/~M31 analogues. If this does have a noticeable effect, we expect that the surface brightness profile features due to apocentre pile-ups will be more concentrated than suggested by our simulations. On the other hand, the stars defining the ``stellar splashback'' primarily come about from the extended stellar haloes of accreted massive dwarf galaxies, rather than their discs (see e.g. the origin of the 2-pericentre particles on the left of Fig.~\ref{fig9}, or the binding energy distribution in Fig.~\ref{fig10}), so as long as these dwarf halo stars exist we expect that the stellar halo ``edge'', as defined by the steepest drop in the log-slope of the stellar density profile, will not change significantly. What may change is the surface brightness at which the features are observed. The work of \citet{keller} describes in detail how the numerical implementation of supernovae feedback in simulations can alter the morphology of dwarf galaxies, leading to brighter or fainter stellar haloes of Milky Way-mass galaxies. 

We have shown that the stripping of dwarf galaxy stars leads to the
formation of shells in phase space that traces similar shells in the
dark matter distribution. It would be interesting to explore whether
these features are affected by the nature of the dark matter. For
example, warm dark matter leads to the suppression of the matter power
spectrum on small scales, such that low-mass haloes and the fraction
of the smooth component made up of unresolved haloes would not exist
in our simulations, potentially leading to more distinctive caustic
features in the dark matter, and perhaps also in the stars, if small
subhaloes perturb streams of stripped particles in CDM 
\citep{lovell_wdm_sfr}.

Self-interacting dark matter (SIDM) may also result in a different
phase-space picture. Firstly, dark matter haloes may suffer enhanced
disruption due to `dark ram pressure' stripping from the host galaxy
halo \citep{selfintdm,sirks}. In some SIDM models cores form in dark
matter haloes. Core formation is also expected in CDM galaxy formation
models in which baryonic feedback is sufficiently impulsive
\cite{benitez-llambay}. It has been shown that tidal stripping is more
efficient in haloes with cores, enhancing subhalo disruption and
producing wider stellar and dark matter streams
\citep{errani_streams}. This would lead to fuzzier streams of stripped
particles and smoother halo density profiles, in which changes in the
smoothly declining density slope are less easily identifiable.


\section*{Acknowledgements}

AD and CSF are supported by the Science and Technology Facilities
Council (STFC) [grant number ST/F001166/1, ST/I00162X/1, ST/P000541/1]. AD is supported by a Royal Society
University Research Fellowship. CSF acknowledges aEuropean Research
Council (ERC) Advanced Investigator grant DMIDAS (GA 786910). This
work used the DiRAC Data Centric system at Durham University, operated
by the ICC on behalf of the STFC DiRAC HPC Facility
(www.dirac.ac.uk). This equipment was funded by BIS National
E-infrastructure capital grant ST/K00042X/1, STFC capital grant
ST/H008519/1, and STFC DiRAC Operations grant ST/K003267/1 and Durham
University. DiRAC is part of the National E-Infrastructure.

\section*{Data availability}
The data analysed in this article can be made available upon reasonable 
request to the corresponding author.




\bibliographystyle{mnras}
\bibliography{edge.bib} 





\bsp	
\label{lastpage}
\end{document}